\documentclass{article}
\usepackage{amssymb,graphicx,cancel}
\usepackage[margin=2cm]{geometry}
\usepackage{slashed}
\usepackage{hepunits}
\usepackage{color}
\usepackage[table]{xcolor}
\usepackage{graphicx}

 \usepackage{jheppub}
\begin{document}
\newcommand{\be}{\begin{eqnarray}}
\newcommand{\ee}{\end{eqnarray}}
\newcommand{\Gev}{\,\,\mathrm{GeV}}
\newcommand{\SUWeak}{\mathrm{SU}(2)_{\rm W}}
\newcommand{\Lag}{\mathcal{L}}
\newcommand{\Lagtree}{\mathcal{L}_{\rm tree}}
\newcommand{\benum}{\begin{enumerate}}
\newcommand{\eenum}{\end{enumerate}}
\newcommand{\bi}{\begin{itemize}}
\newcommand{\ei}{\end{itemize}}
\newcommand{\met}{\slashed{E_T}}
\newcommand{\apr}{{A^\prime}}
\newcommand{\zp}{{Z^\prime}}
\newcommand{\neff}{{N_{\rm eff}}}
\newcommand{\zpr}{{Z^\prime}}
\newcommand{\gmt}{{g_{\mu-\tau}}}

\def\lsim{\mathrel{\raise.3ex\hbox{$<$\kern-.75em\lower1ex\hbox{$\sim$}}}}
\def\gsim{\mathrel{\raise.3ex\hbox{$>$\kern-.75em\lower1ex\hbox{$\sim$}}}}

\preprint{\\ FERMILAB-PUB-19-001-A, LPT-Orsay-18-15, IFIC-19-02, KCL-19-01, IFT-UAM/CSIC-19-7}

\newcommand\FNAL{Fermi National Accelerator Laboratory,  Batavia, IL USA}
\newcommand\Valencia{
Instituto de F\'{\i}sica Corpuscular (IFIC)$,$ CSIC-Universitat de Val\`encia, Valencia, Spain
}
\newcommand\Kings{From 09/18: Department of Physics, King's College London, Strand, London WC2R 2LS, UK}
\newcommand\UCastro{University of Chicago, Department of Astronomy and Astrophysics, Chicago, IL, USA}
\newcommand\KICP{University of Chicago, Kavli Institute for Cosmological Physics, Chicago, IL USA}
\newcommand\Orsay{Laboratoire de Physique Th\'{e}orique (UMR8627), CNRS, Univ. Paris-Sud, Universit\'{e} Paris-Saclay, 91405 Orsay, France}
\newcommand\IFT{From 10/18: Instituto de F\'{i}sica Te\'{o}rica (IFT) UAM-CSIC, Campus de Cantoblanco, 28049 Madrid, Spain}
\newcommand\UAM{From 10/18: Departamento de F\'{i}sica Te\'{o}rica, Universidad Autónoma de Madrid (UAM), Campus de
Cantoblanco, 28049 Madrid, Spain}

\newcommand{\gk}[1]{\textbf{\textcolor{red}{(#1 --gk)}}}
\newcommand{\ME}[1]{\textbf{\textcolor{blue}{(#1 --ME)}}}
\newcommand{\MPIERRE}[1]{\textcolor{red}{\textbf{[MP: #1]}}}

\author[a,b]{Miguel Escudero,}
\author[c,d,e]{Dan Hooper,}
\author[c]{Gordan Krnjaic,}
\author[f,g,h]{Mathias Pierre}

\affiliation[a]{\Kings}
\affiliation[b]{\Valencia}
\affiliation[c]{\FNAL}
\affiliation[d]{\UCastro}
\affiliation[e]{\KICP}
\affiliation[f]{\Orsay}
\affiliation[g]{\IFT}
\affiliation[h]{\UAM}

\title{Cosmology With a Very Light $L_\mu - L_\tau$ Gauge Boson}

\abstract{In this paper, we explore in detail the cosmological implications of an abelian $L_\mu-L_\tau$ gauge extension of the Standard Model featuring a light and weakly coupled $Z'$. Such a scenario is motivated by the longstanding $\sim$\,$4 \sigma$ discrepancy between the measured and predicted values of the muon's anomalous magnetic moment, $(g-2)_\mu$, as well as the tension between late and early time determinations of the Hubble constant. If sufficiently light, the $Z'$ population will decay to neutrinos, increasing the overall energy density of radiation and altering the expansion history of the early universe. We identify two distinct regions of parameter space in this model in which the Hubble tension can be significantly relaxed. The first of these is the previously identified region in which a $\sim$\,$10-20$ MeV $Z'$ reaches equilibrium in the early universe and then decays, heating the neutrino population and delaying the process of neutrino decoupling. For a coupling of $g_{\mu-\tau} \simeq (3-8) \times 10^{-4}$, such a particle can also explain the observed $(g-2)_{\mu}$ anomaly. 
In the second region, the $Z'$ is very light ($m_{Z'} \sim 1\,\text{eV}$ to $\text{MeV}$) and very weakly coupled ($g_{\mu-\tau} \sim 10^{-13}$ to $10^{-9}$). In this case, the $Z'$ population is produced through freeze-in, and decays to neutrinos after neutrino decoupling. Across large regions of parameter space, we predict a contribution to the energy density of radiation that can appreciably relax the reported Hubble tension, $\Delta N_{\rm eff} \simeq 0.2$.}

\emailAdd{miguel.escudero@kcl.ac.uk}
\emailAdd{dhooper@fnal.gov}
\emailAdd{krnjaicg@fnal.gov}
\emailAdd{mathias.pierre@uam.es}
\maketitle

\section{Introduction} 
\label{sec:intro}

Light and weakly coupled particles are found within many well-motivated extensions of the Standard Model (SM)~\cite{Essig:2013lka,Alexander:2016aln}. Among the motivations for such states is the muon's anomalous magnetic moment, whose measured value constitutes a $4.1 \sigma$ discrepancy with respect to the predictions of the SM~\cite{Tanabashi:2018oca}. This anomaly has inspired many explanations involving new sub-GeV particles~\cite{Pospelov:2008zw,Altmannshofer:2014pba}. Although many of these explanations have already been ruled out (including both visibly and invisibly decaying dark photons with kinetic mixing $\epsilon \sim 10^{-3}$), scenarios in which new states couple predominantly to muons remain viable.

In addition to the muon's magnetic moment, there is also a $\sim$\,3$\sigma$ tension between the values of the Hubble constant, $H_0$, as determined from local measurements~\cite{Riess:2018byc,Riess:2016jrr} and from the temperature anisotropies of the cosmic microwave background (CMB)~\cite{Aghanim:2018eyx}. Such a discrepancy could be ameliorated if the expansion rate of the universe departed from the predictions of standard $\Lambda$CDM cosmology at early times~\cite{Bernal:2016gxb,Aylor:2018drw}. Particularly well motivated within this context are scenarios in which the energy density in relativistic particles exceeds that predicted by the SM, generally parameterized in terms of a non-zero contribution to the effective number of neutrino species, $\Delta N_{\rm eff}$~\cite{Weinberg:2013kea,Shakya:2016oxf,Berlin:2018ztp,DEramo:2018vss,Dessert:2018khu}. Scenarios involving early dark energy~\cite{Poulin:2018cxd,Poulin:2018dzj,Poulin:2018zxs} or a component of decaying dark matter~\cite{Bringmann:2018jpr} have also been proposed to address this tension. Upcoming CMB measurements will be significantly more sensitive to the value of $\Delta N_{\rm eff}$~\cite{Abazajian:2016yjj}, providing us with further motivation to consider this class of scenarios. 

Particularly interesting with this context are models with a broken $U(1)_{L_\mu - L_\tau}$ gauge symmetry, corresponding to a new massive gauge boson that couples to muons, taus, and their corresponding neutrinos. This is one of the few anomaly free $U(1)$ gauge extensions of the SM~\cite{He:1990pn,He:1991qd}, and is the only such model without tree-level couplings to first generation quarks and/or leptons.\footnote{Other choices for gauging global SM quantum numbers are $B-L, L_i - L_j, B- 3 L_i$, where $B$ and $L$ are respectively baryon and lepton numbers and a subscript denotes a specific lepton flavor.} For this reason, $U(1)_{L_\mu - L_\tau}$ models are relatively unconstrained and lead to qualitatively different phenomenological and cosmological consequences~\cite{Altmannshofer:2014pba,Altmannshofer:2014cfa,Kamada:2015era,Kamada:2018zxi,Foldenauer:2018zrz,Asai:2018ocx,Carena:2018cjh,Arcadi:2018tly,Bauer:2018egk,Altmannshofer:2016jzy,Biswas:2016yan,DiFranzo:2015qea,Bauer:2018onh,Ilten:2018crw,Asai:2018ocx}.

In this paper, we revisit in detail the impact of a $L_{\mu}-L_{\tau}$ gauge boson on the particle content and expansion rate of the early universe. We solve the full set of Boltzmann equations that describe the evolution of the $Z'$ and neutrino populations, calculating the value of $\Delta N_{\rm eff}$ across the parameter space of this model. We identify two distinct regions in which the tension between early and late-time determinations of the Hubble constant can be substantially relaxed. In the first of these regions, a $\sim$\,$10-20$ MeV $\zp$ reaches equilibrium in the early universe and then decays, heating the neutrino population and delaying the process of neutrino decoupling. For a gauge coupling of $g_{\mu-\tau} \simeq (3-8) \times 10^{-4}$, such a $\zp$ is also capable of explaining the measured value of the muon's anomalous magnetic moment. In the second region, a very light and weakly coupled $\zp$ is produced in the early universe through freeze-in, and decays to neutrinos only after neutrino decoupling. We identify a large plateau of parameter space in which the contribution to the energy density is near $\Delta N_{\rm eff} \simeq 0.2$, in good agreement with the value required to relax the tension associated with the Hubble constant.

The remainder of this paper is organized as follows. In Sec.~\ref{sec:model} we describe the $U(1)_{L_\mu - L_\tau}$ model, including kinetic mixing between the $\zp$ and the photon. In Sec.~\ref{sec:early-universe} we solve the full system of Boltzmann equations in order to calculate the contribution to $N_{\rm eff}$ in this model, in both the thermalized and freeze-in regimes. In Sec.~\ref{sec:constraints} we review the experimental and astrophysical constraints on this model. Finally, in Sec.~\ref{sec:conclusion} we summarize our results and conclusions. We also include a series of appendices in which we describe the Boltzmann equations for $\zp$ production, the bounds from Supernova 1987A, the contribution to $\zp$-photon kinetic mixing induced by additional scalars, the $\zp$ mediated corrections to the energy transfer rates responsible for delaying the process of neutrino decoupling, and a detailed description of the solutions to the Boltzmann equation in the freeze-in regime.

\section{Model Description}
\label{sec:model}


In this study, we extend the SM to include a spontaneously broken $U(1)_{L_\mu-L_\tau}$ gauge symmetry. In the broken phase\footnote{The scalar field whose vacuum expectation value is responsible for the spontaneous breaking of $U(1)_{L_\mu-L_\tau}$ is generally expected to be much heavier than the $Z^\prime$ mass and other energy scales considered in this work. Details regarding this symmetry breaking are beyond the scope of this paper and are not expected to affect our results.}, the Lagrangian is given by
\be \label{eq:lag-min}
{\cal L} = {\cal L}_{\rm SM} - \frac{1}{4} {Z^\prime}^{\alpha\beta}Z^\prime_{\alpha\beta} + \frac{m^2_\zp}{2}  {Z^\prime_\alpha}  {Z^\prime}^\alpha    + Z^{\prime}_\alpha J^{\alpha}_{\mu - \tau}, ~~
\ee
where $g_{\mu-\tau}$ is the gauge coupling, $m_\zp$ is the mass of the gauge boson, and $ Z^\prime_{\alpha\beta} \equiv \partial_\alpha Z^\prime_\beta - \partial_\beta Z^\prime_\alpha$ 
is the field strength tensor. The $\mu -\tau$ current is 
\be \label{eq:jmutau}
J^{\alpha}_{\mu - \tau} = g_{\mu - \tau} \left(    \bar \mu \gamma^\alpha \mu + \bar \nu_{\mu} \gamma^\alpha P_L \nu_{\mu}  -
  \bar \tau \gamma^\alpha \tau -  \bar \nu_{\tau} \gamma^\alpha P_L \nu_{\tau}    \right),
\ee
where $P_L = \frac{1}{2}(1-\gamma_5)$ is the left
chirality projector and
\be
\Gamma_{\zp \to  \ell^+ \ell^-} =  \frac{g^2_{\mu-\tau} m_{\zp}}{12 \pi } \left( 1 + \frac{2m^2_\ell}{m^2_{\zp}} \right)     \sqrt{1 - \frac{4 m_\ell^2}{m^2_{\zp} }}, ~~~~~~~~~~~~~~
\Gamma_{\zp \to \bar \nu_i\nu_i} = \frac{g^2_{\mu-\tau} m_{\zp}}{24 \pi} ,~~~
\ee
are the rest frame partial widths for $\zpr$ decays to charged leptons $\ell =\mu, \tau$ and neutrinos $\nu_i = \nu_{\mu, \tau}$.

The contribution of the $\zp$ to the anomalous magnetic moment of the muon, to leading order in $g_{\mu-\tau}$, is given by~\cite{Pospelov:2008zw}
\be
\Delta a^{Z^\prime}_{\mu} = \frac{g^2_{\mu-\tau} }{8\pi^2} \int_0^1 d z \frac{ 2m^2_\mu z (1-z)^2}{ m_\mu^2 (1-z)^2 +  m_{Z^\prime}^2 \, z} \simeq  1.3 \times 10^{-10} \, \left( \frac{\gmt}{10^{-4}} \right)^2  ,
\ee
where $a_{\mu}  \equiv \frac{1}{2}(g-2)_\mu$. In the last step, we have taken the  $m_{\zp} \ll m_\mu$ limit. The measured value of this quantity is $\Delta a_\mu \equiv a_\mu({\rm obs})  - a_\mu({\rm SM}) = (28.8 \pm 8.0) \times 10^{-10}$~\cite{Tanabashi:2018oca}, and thus requires a gauge coupling of $g_{\mu-\tau} \simeq (3-8) \times 10^{-4}$ in order to resolve the anomaly.

\begin{figure}[t]
\centering
\begin{tabular}{cc}
\includegraphics[width=0.4\textwidth]{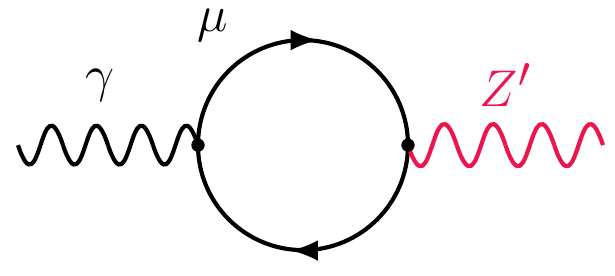} & \includegraphics[width=0.4\textwidth]{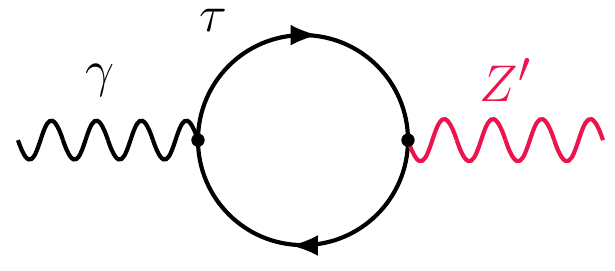} 
\end{tabular}
\caption{Feynman diagrams that lead to kinetic mixing between the $\zp$ and the photon. In addition to the diagrams shown here, there could also be model dependent contributions that arise from heavy states charged under both electromagnetism and $U(1)_{L_{\mu}-L_{\tau}}$. For a treatment of additional contributions from 
heavy physics beyond the SM, see Appendix~\ref{app:mu-tau_loop}.}\label{fig:diagram}
\end{figure}

At tree-level, the $\zp$ in this model couples only to heavy leptons and their neutrino flavors. Muon and tau loops, however, can lead to kinetic mixing between the $\zp$ and the photon, inducing an effective coupling for the $\zp$ to electromagnetically charged fermions. For low energy processes we can integrate out $\mu$ and $\tau$, resulting in an off-diagonal kinetic term, $F^{\alpha\beta} Z^\prime_{\alpha\beta}$, between the $\zp$ and the photon. Diagonalizing these fields and restoring canonical normalization induces the following $\zp$ coupling to the electromagnetic current
\be \label{eq:jem}
{\cal L} \supset Z^\prime_\alpha \left( J_{\mu-\tau}^\alpha  + \epsilon J^\alpha_{\rm EM}\right), ~~~~~~ J^\alpha_{\rm EM} = e \sum_f Q_f \bar f \gamma^\alpha f,
\ee
where $e$ is the electron charge, $f$ is a SM fermion with charge $Q_f$, and the quantity $\epsilon$ quantifies the degree of kinetic mixing. The irreducible contributions to $\epsilon$ from the loops shown in Fig.~\ref{fig:diagram} are given by~\cite{Kamada:2015era}
\be\label{eq:kinmix}
\epsilon =  -\frac{e g_{\mu-\tau}}{2\pi^2}  \int^1_0 d x \,  x (1-x) \log \left[\frac{m_\tau^{2} - x(1-x)q^2}{m_\mu^{2} - x(1-x)q^2}\right]  
~ \xrightarrow[  m_\mu \gg q]{} ~ 
- \frac{e g_{\mu-\tau}}{12 \pi^2}  \log \frac{m_\tau^2}{m_\mu^2}  
\simeq -  \frac{      g_{\mu -\tau} }{70}.
 \label{eq:g-Zp_mixing}
\ee
This calculation provides us with a benchmark value for $\epsilon$, which we will refer to throughout this paper as ``natural kinetic mixing''. One should keep in mind, however, that other model dependent contributions could potentially arise, in particular if there exist heavy particles that are charged under both electromagnentism and $L_{\mu}-L_{\tau}$.

As a result of this kinetic mixing, the $\zp$ can decay to $e^+ e^-$ with a partial width given by
 \be
 \label{eq-zp-decay-electrons}
 \Gamma_{\zp \to  e^+ e^-} =  \frac{(\epsilon e)^2  m_{\zp} }{12 \pi } \left( 1 + \frac{2m^2_e}{\>m^2_{\zp}} \right)     \sqrt{1 - \frac{4 m_e^2}{\> m^2_{\zp} }},
 \ee
leading to the following branching fraction
 \be
 \label{eq:branching}
 {\rm Br}_{Z'\to e^+e^-}  = \frac{ \Gamma_{\zp \to e^+ e^-}  }{  \Gamma_{\zp \to \bar \nu_\mu \nu_\mu } + \Gamma_{\zp \to \bar \nu_\tau \nu_\tau }}
 \simeq   \left( \frac{e\epsilon }{\, \, g_{\mu - \tau}} \right)^2 \simeq  2 \times 10^{-5},
 \ee
where in the last step we have used the value of $\epsilon$ given in Eq.~(\ref{eq:g-Zp_mixing}) and taken the 
 $m_e \ll  m_\zp \ll m_\mu$ limit. Note that this result is independent of the gauge coupling,
which cancels in the absence of additional contributions to $\epsilon$. 

\section{Contributions to $N_{\rm eff}$ }
\label{sec:early-universe}

In the early universe, the $\zp$ number density, $n_{\zp}$, is governed by the following Boltzmann equation 
\be \label{eq:zp-boltz}
\dot n_\zp + 3 H n_{\zp}  =   \langle \Gamma_{\zpr}  \rangle   \left(  n_{\zpr}^{\rm (eq)}  -    n_{\zpr} \right),
\ee
where $H \equiv \dot{a}/a$ is the expansion rate of the universe and $n^{\rm (eq)}_{\zp}$ is the equilibrium value of the $\zp$ number density. The quantity $\Gamma_{\zp}$ is the rest frame width, for which the thermally averaged value is given by the following
\be 
\langle \Gamma_{\zp} \rangle  \equiv \Gamma_{\zp}  \frac{ K_1(x) }{  K_2(x) },
\ee
where $K_{1,2}$ are Bessel functions of the first and second kind and $x \equiv m_{\zp}/T$. Although many processes can affect $n_{\zp}$, in the weakly coupled regime ($g_{\mu-\tau} \ll 1$) it suffices to consider only decays and inverse decays in the collision term. For a derivation and other details, see Appendix~\ref{app:Boltzmann_Zp}.

We  are interested in the effect of $\zp$ decays on the  total radiation density just prior to matter-radiation equality at $T_\gamma \simeq 0.8\, \eV$, which can be written in terms of $N_{\rm eff}$, the effective number of neutrino species:  
\be
\rho_R = \left[    1 +  \frac{7}{8} \left(\frac{4}{11} \right)^{4/3}  N_{\rm eff}  \right] \rho_\gamma,
\ee
where $\rho_\gamma$  is the photon energy density, the factor of $7/8$ accounts for the fact that neutrinos are fermions, and the $(4/11)^{1/3} = T_{\nu}/T_\gamma$ in the SM. Note that the SM prediction for $N_{\rm eff}^{\rm SM}  = 3.045$~\cite{deSalas:2016ztq,Mangano:2005cc} is slightly larger than 3 because of the entropy transferred to the neutrinos through $e^+e^-$ annihilations, the non-instantaneous nature of neutrino decoupling, finite temperature corrections, and neutrino oscillations~\cite{deSalas:2016ztq,Mangano:2005cc,Dolgov:1997mb,Dicus:1982bz}.

\begin{figure}[t]
\centering
\includegraphics[width=1\textwidth]{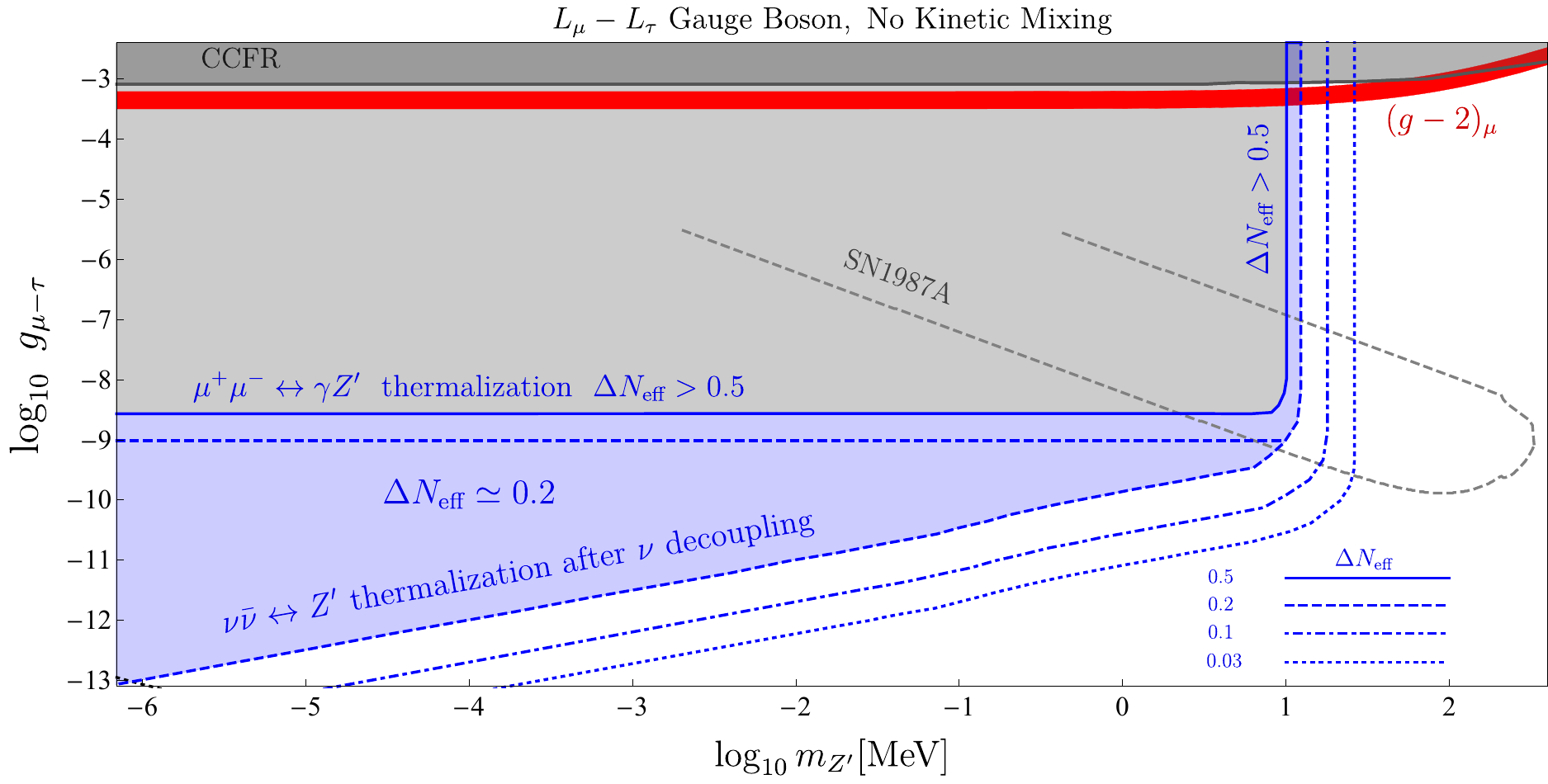} 
\includegraphics[width=1\textwidth]{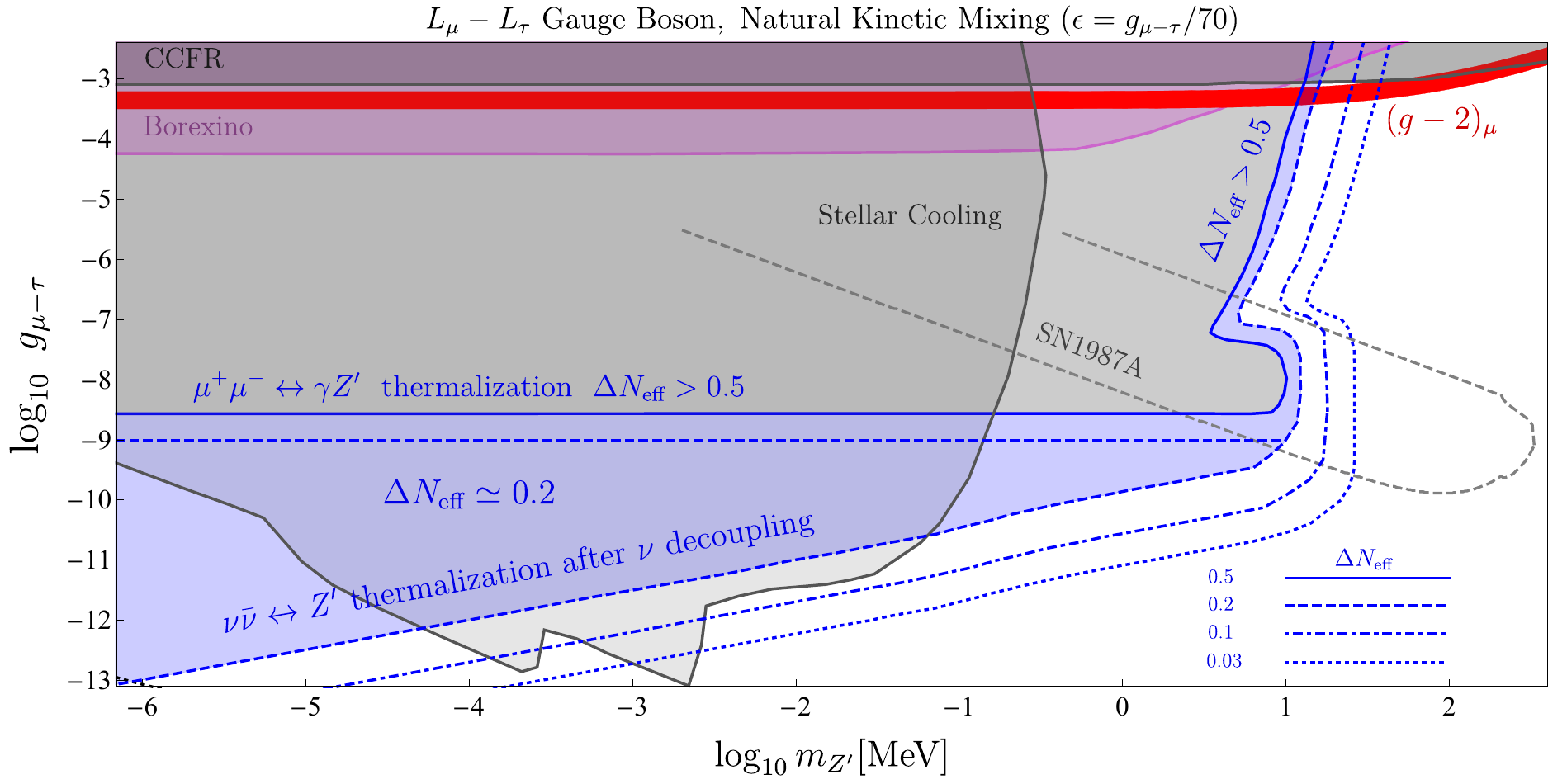} 
\caption{A summary of the phenomenology of a $L_\mu - L_\tau$ gauge boson in the absence of kinetic mixing ($\epsilon=0$) or assuming the ``natural'' degree of kinetic mixing associated with the diagrams shown in Fig.~\ref{fig:diagram} ($\epsilon \simeq g_{\mu-\tau}/70$). We show contours of constant $\Delta N_\text{eff}$, as well as the regions favored by measurements of the muon's anomalous magnetic moment, $(g-2)_\mu$ (red band). We also identify the regions that are excluded by measurements of di-muon production in neutrino nucleus scattering (CCFR)~\cite{Mishra:1991bv,Altmannshofer:2014pba}, solar neutrino observations (Borexino)~\cite{Harnik:2012ni,Laha:2013xua,Kamada:2015era}, by stellar cooling~\cite{An:2013yfc,Hardy:2016kme} and by observations of Supernova 1987A~\cite{Kolb:1987qy} (see Sec.~\ref{sec:constraints} and Appendix~\ref{app:SNbounds}). Regions shaded in blue yield $\Delta N_{\rm eff} = 0.2-0.5$ which can substantially ameliorate the tension between early and late time measurements of $H_0$.}
\label{fig:money}
\end{figure}

The evolution of the $\zp$ population in the early universe depends on the values of its mass and coupling. Broadly speaking, we will consider two qualitatively distinct regions of parameter space:

\begin{itemize}

\item{\bf Early Universe Equilibrium:} If $g_{\mu-\tau} \gtrsim
 4 \times 10^{-9}$, the $Z^\prime$ population thermalizes with the SM bath at early times and decays into neutrinos when $T \sim m_{Z^\prime}/3$. 
 If these decays occur predominantly after the neutrinos and photons decouple, they contribute to the neutrino energy density and thereby increase the value of $N_{\rm eff}$.
 Furthermore, in the presence of non-negligible kinetic mixing with the photon, $\zp$ interactions with charged particles can delay the neutrino-photon decoupling, quantitatively affecting $N_{\rm eff}$ .
  
  \item{\bf Freeze In (Late Equilibration):} If $g_{\mu-\tau} \lesssim 4 \times 10^{-9}$, the $\zp$ population will not have initially been in equilibrium with the SM in the very early universe, but is instead produced through the freeze-in mechanism. For a wide range of masses and couplings, the $\zp$ production rate is slower than Hubble expansion at very
  early times, but then becomes comparable as the Hubble rate decreases. Across this broad region of parameter space, the $\zp$ population eventually thermalizes with neutrinos, but only {\it after} the latter decouple from
  photons, inducing a contribution of $\Delta N_{\rm eff} \simeq 0.21$ through $\zp\to \bar \nu \nu$ decays, provided that $m_{Z'} \gtrsim  1\,\text{eV}$ so that the $Z'$ decays prior to CMB formation.  
 
\end{itemize}

Each of these parameter space regions can be easily identified in Fig.~\ref{fig:money}. 

\begin{figure}[t]
\centering
\begin{tabular}{cccc}
&   \includegraphics[height=0.15\textwidth]{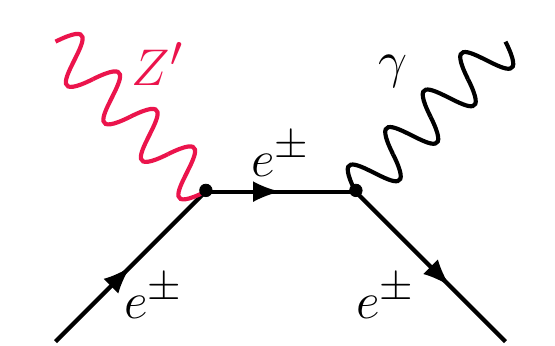} &  \includegraphics[height=0.15\textwidth]{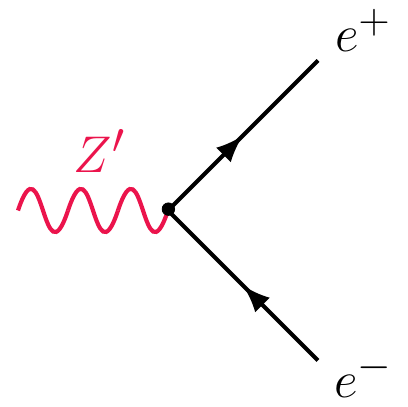} & \\
 \includegraphics[height=0.2\textwidth]{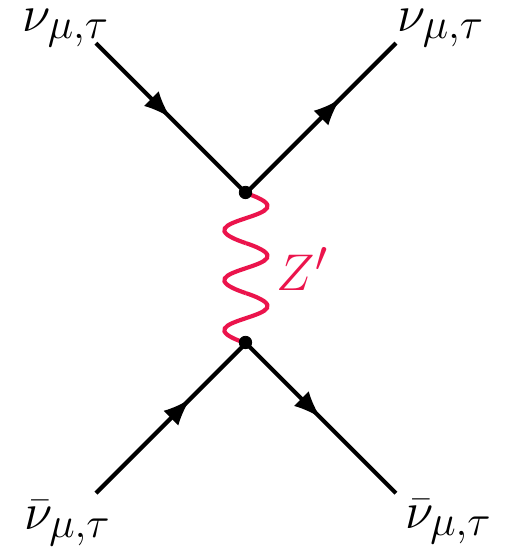} & \includegraphics[height=0.15\textwidth]{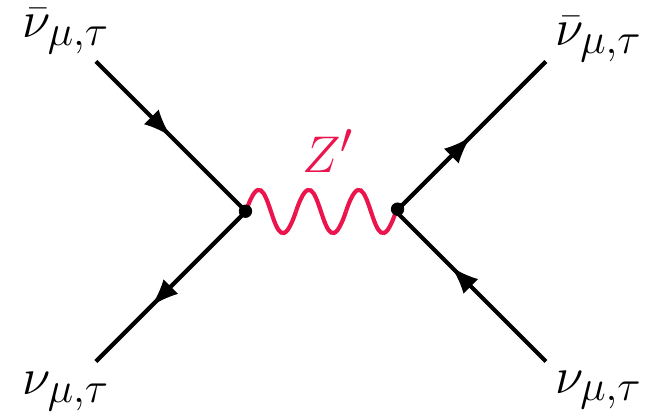} & \includegraphics[height=0.15\textwidth]{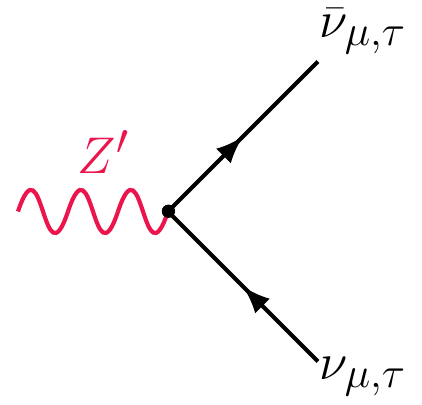} & \includegraphics[height=0.2\textwidth]{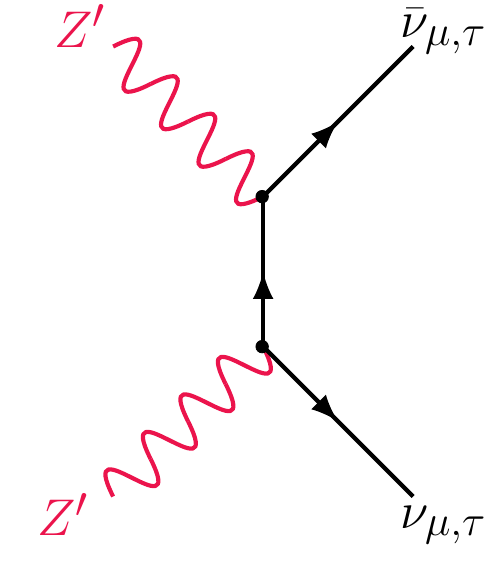} 
\end{tabular}
\caption{$\zp$ induced scattering and decay processes that can alter the process of $\nu_{\mu}$ and $\nu_{\tau}$ decoupling. }\label{fig:diagrams}
\end{figure}

\subsection{Early Universe Equilibrium Regime}
\label{sec:equilibrium-regime}
If $g_{\mu-\tau}$ is sufficiently large, the rate of $\zp$ production will exceed that of Hubble expansion, $\langle \Gamma_\zp  \rangle \gg H$, keeping the $\zp$ population in equilibrium with the SM plasma at early times. In this limit, the $\zp$ population satisfies $n_{\zp} = n_{\zp}^{\rm (eq)}$, where 
 \be
 n^{\rm (eq)}_{\zp} =  \int_0^\infty \frac{d^3  \vec  p}{(2\pi)^3} \frac{g_\zp }{e^{E/T} - 1},
 \ee
 is the equilibrium number density and $g_\zp  = 3$ is the number of spin states. As the temperature of the universe drops below $m_{\zp}$, inverse decays become kinetically forbidden and the entropy of this population is transferred to other species. If only neutrinos and photons are present, we can write the effective number of neutrino species as  
 \be
 \label{eq:neff-eq-noeps}
 N_{\rm eff} = \frac{8}{7} \left( \frac{11}{4} \right)^{4/3}  \frac{ \rho_{\nu}  }{ \rho_\gamma}     \, ,
 \ee
 where $\rho_\nu$ generically differs from its SM value due to entropy transferred from $\zp$ decays. 
In the equilibrium regime, these decays take place while $\zp$ mediated 
$\nu$ scattering is in chemical equilibrium, so the decay daughter neutrinos 
thermalize with the existing neutrino population and increase their common temperature. \\

The presence of a light $U(1)_{L_{\mu}-L_{\tau}}$ $Z^\prime$ can substantially alter the decoupling history of the neutrinos through processes of the kind shown in Fig.~\ref{fig:diagrams}. Of these diagrams, the most important are the decays and inverse decays of the $\zp$ to neutrinos and electrons. Observations from Borexino constrain the scattering rate between neutrinos and electrons to be within 8\% of the SM prediction~\cite{Harnik:2012ni,Kamada:2015era}, and thus such $2\rightarrow 2$ processes cannot significantly impact the process of neutrino decoupling.
The impact of $\nu_\mu \bar{\nu}_\tau \to Z' Z'$ scattering is to suppress any chemical potential that might exist in the $\nu_\mu - \nu_\tau - Z'$ sector, an effect that will be efficient for $g_{\mu- \tau }\gtrsim 10^{-5}$~\cite{Kamada:2015era}. Throughout the phenomenologically viable parameter space of this model (with $N_\text{eff} \lsim 4$), however, the chemical potential will be negligible as a result of $e^+e^- \leftrightarrow \nu_{\mu,\,\tau} \bar{\nu}_{\mu,\,\tau}$ scattering, which is efficient for $T \gsim 3\,\text{MeV}$. 
Finally, the process $e^+e^- \leftrightarrow Z' \gamma$ is suppressed by a factor of $\alpha \sim 1/137$ relative to $e^+e^- \leftrightarrow Z'$ and thus can be safely neglected. 

From these considerations, we conclude that it is sufficient to calculate the evolution of the following three temperatures: $T_\gamma (=$\,$T_e), \, T_{\nu_e}$ and $T_{\nu_\mu} (=$\,$T_{\nu_\tau}$\,$=$\,$T_{Z'})$. We adopt the assumption of Maxwell-Boltzmann statistics in the collision terms to derive the following set of coupled differential equations which describe the evolution of these quantities~\cite{Escudero:2018mvt}:
\be
\frac{dT_{\nu_e}}{dt} &=& -           \left(4 H \rho_{\nu_e}  - \frac{                      \delta \rho_{\nu_e}        }{\delta t}   \right)      \left( \frac{   \partial \rho_{\nu_e} }{\partial T_{\nu_e} }   \right)^{-1}         \,  \label{eq:Tnue_mu-tau}  \\
\frac{dT_{\nu_\mu}}{dt} &=& -  \left( 8 H  \rho_{\nu_\mu} + 3 H \left( \rho_{Z'} + p_{Z'}\right)  - 2 \frac{ \delta \rho_{\nu_\mu}}{\delta t}-  \frac{ \delta \rho_{Z'}}{\delta t}    \right) \left(  2 \frac{\partial \rho_{\nu_\mu} }{ \partial T_{\nu_\mu} } + \frac{\partial \rho_{Z'}}{ \partial T_{\nu_\mu}  }  \right)^{-1}   \, \label{eq:Tnumu_mu-tau} \\
\frac{dT_{\gamma}}{dt}  &=&- \left(  4 H \rho_{\gamma} + 3 H \left( \rho_{e} + p_{e}\right) + \frac{ \delta \rho_{\nu_e}}{\delta t} + 2  \frac{ \delta \rho_{\nu_\mu}}{\delta t} + \frac{ \delta \rho_{Z'}}{\delta t}  \right)\left(    \frac{\partial \rho_{\gamma}}{\partial T_\gamma} + \frac{\partial \rho_e}{\partial T_\gamma}       \right)^{-1}    ,
\label{eq:Tgam_mu-tau}
\ee
where the energy transfer rates are
\be
\frac{ \delta \rho_{Z'}}{\delta t} & =&  \frac{3 m_{Z'}^3}{2\pi^2} \left[ T_{\gamma} K_2\left(\frac{m_{Z'}}{T_{\gamma}}\right) - T_{\nu_\mu} K_2\left(\frac{m_{Z'}}{T_{\nu_\mu}}\right) \right]   \Gamma_{Z' \to e^+ e^-} \label{eq:DrhoZp}  \\
\frac{ \delta \rho_{\nu_e}}{\delta t}&=& \frac{G_F^2}{\pi^5} \biggl[   2( 1 + 4 s_W^2 + 8 s_W^4 ) F(T_\gamma, T_{\nu_e}) + F(T_{\nu_\mu}, T_{\nu_e})     \biggr] \label{eq:Drhonue}  \\
\frac{ \delta \rho_{\nu_\mu}}{\delta t} &=&    
\frac{G_F^2}{\pi^5} \biggl[   2( 1 - 4 s_W^2 + 8 s_W^4 ) F(T_\gamma, T_{\nu_\mu}) - \frac{1}{2} F(T_{\nu_\mu}, T_{\nu_e})     \biggr] + 
\frac{  e^2\epsilon^2 g_{\mu-\tau}^2} {4 \pi^5 m_{Z'}^4} F(T_\gamma, T_{\nu_\mu}) \label{eq:Drhonumu} ,
\ee
where $G_F = 1.16 \times 10^{-5} \, \GeV^{-2}$ is the Fermi constant, $s_W^2 = 0.223$ is the square of the weak mixing angle~\cite{Tanabashi:2018oca}, $K_2$ is a modified Bessel function of the second kind and we have 
defined the function
\be
F(a,b) \equiv 16(a^9 - b^9) + 7a^4 b^4 (a-b),
\ee
where the first and second terms arise from annihilation and scattering processes respectively; for a derivation of the $\zp$ induced 
corrections, see Appendix~\ref{app:App_nudec}.
 In the above equations, the $\delta \rho_\nu$ terms account for SM and $\zp$ mediated neutrino-electron and neutrino-neutrino interactions, whereas the $\delta \rho_{Z'}$ term accounts for $e^+e^- \leftrightarrow Z'$ inverse decays. Here we have safely assumed that the $Z'$ population is strongly coupled to the $\mu-\tau$ neutrinos due to the large decay rate, $\Gamma_{Z'\to \bar{\nu}_{\mu} \nu_\mu} \gg H$. The derivation
 of SM collision terms here can be found in~\cite{Hannestad:1995rs,Kawasaki:2000en,Dolgov:2002ab,Escudero:2018mvt} and the collision
  terms for the $\zp$ mediated processes are derived in Appendix~\ref{app:App_nudec}.

Finally, we note that the mass splittings inferred from atmospheric and solar neutrino observations correspond to oscillations that are active for temperatures $T$\,$\sim 3$ MeV and $T$\,$\sim 5$ MeV, respectively~\cite{Hannestad:2001iy,Dolgov:2002ab,Dolgov:2002wy}. Since neutrino oscillations imply a rapid change in the neutrino flavor, they act as to equilibrate the $\nu_e$ and $\nu_\mu - \nu_\tau$ distribution functions, leading to $T_{\nu_e} \simeq T_{\nu_\mu}$. To properly account for such oscillations would require the use of the density matrix formalism. Since the effect of such oscillations is to equilibrate the temperatures of the different neutrino species, however, we can simply work in terms of a common neutrino temperature, $T_\nu$. In this case, the evolution of the temperatures $T_\nu$ and $T_\gamma$ can be simplified as follows
\be
\frac{dT_{\nu}}{dt} &=& -  \left( 4 H  \rho_{\nu} + 3 H \left( \rho_{Z'} + p_{Z'}\right)  -  \frac{ \delta \rho_{\nu}}{\delta t}-  \frac{ \delta \rho_{Z'}}{\delta t}    \right) \left(   \frac{\partial \rho_{\nu} }{ \partial T_{\nu} } + \frac{\partial \rho_{Z'}}{ \partial T_{\nu}  }  \right)^{-1}   \label{eq:Tnu_mu-tau} \\
\frac{dT_{\gamma}}{dt}  &=&- \left(  4 H \rho_{\gamma} + 3 H \left( \rho_{e} + p_{e}\right) + \frac{ \delta \rho_{\nu}}{\delta t} +  \frac{ \delta \rho_{Z'}}{\delta t}  \right)\left(    \frac{\partial \rho_{\gamma}}{\partial T_\gamma} + \frac{\partial \rho_e}{\partial T_\gamma}       \right)^{-1}, 
\label{eq:Tgam_mu-tau_one}
\ee
where $\rho_\nu \equiv \rho_{\nu_e} + 2\rho_{\nu_\mu} $ and  the energy transfer rates are $\delta \rho_\nu /\delta t = \delta \rho_{\nu_e}/\delta t + 2\delta \rho_{\nu_\mu}/\delta t$ and $\delta \rho_{Z'}/\delta t$, as described in Eqns.~\eqref{eq:DrhoZp},~\eqref{eq:Drhonue} and~\eqref{eq:Drhonumu}.

\begin{figure}[t]
\centering
\begin{tabular}{cc}
\includegraphics[width=0.49\textwidth]{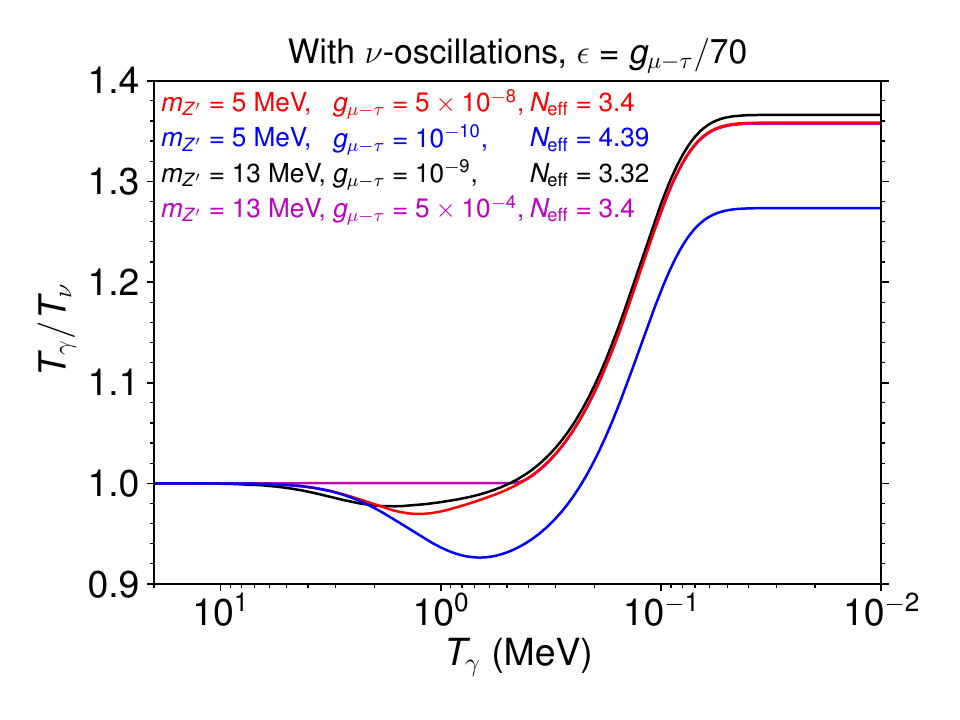} &\includegraphics[width=0.49\textwidth]{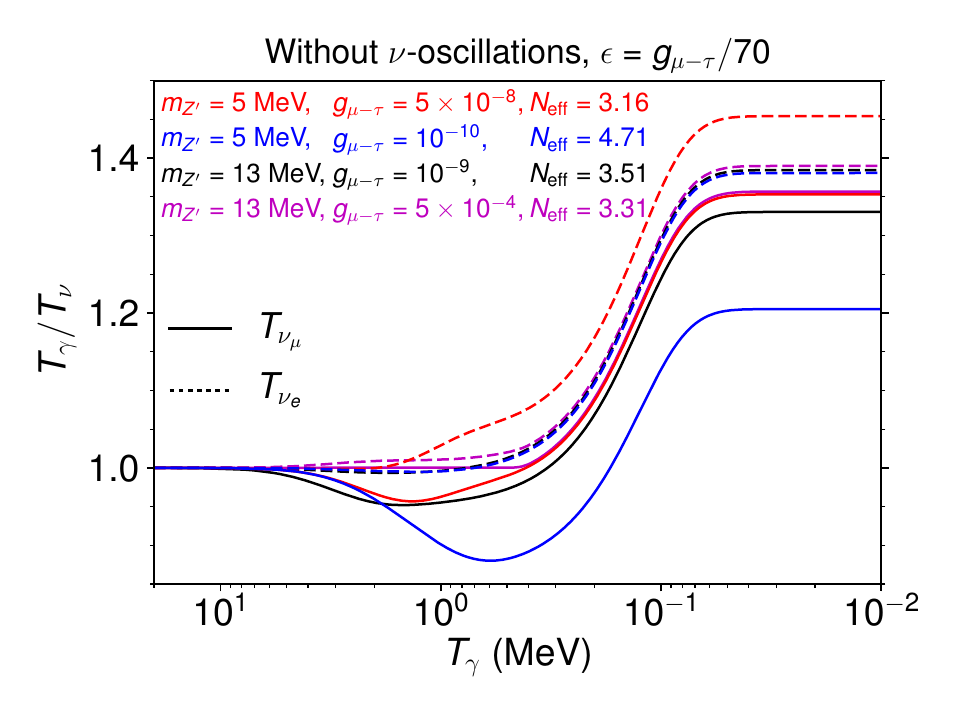}
\end{tabular}
\caption{The evolution of $T_{\gamma}/T_{\nu}$ for several choices of the $\zp$ mass and $L_{\mu}-L_{\tau}$ gauge coupling, adopting the ``natural'' value for the kinetic mixing parameter, $\epsilon \simeq g_{\mu-\tau}/70$. The solid lines denote $T_{\nu_{\mu}, \nu_{\tau}}$ while the dashed lines correspond to $T_{\nu_e}$. In the left panel, we adopt $T_{\nu_e}=T_{\nu_\mu}$ as expected from rapid neutrino oscillations. In the right panel, neutrino oscillations are neglected.}\label{fig:Tplots}
\end{figure}

\begin{figure}[t]
\centering\hspace{-0.8cm}
\begin{tabular}{cc}
\includegraphics[width=0.45\textwidth]{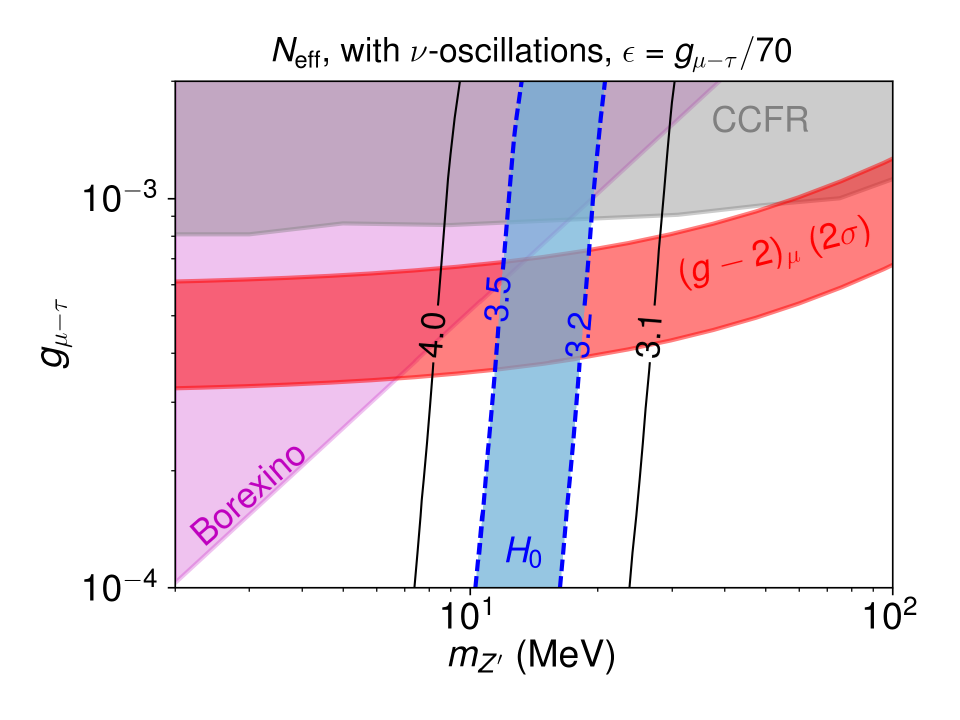} &\includegraphics[width=0.45\textwidth]{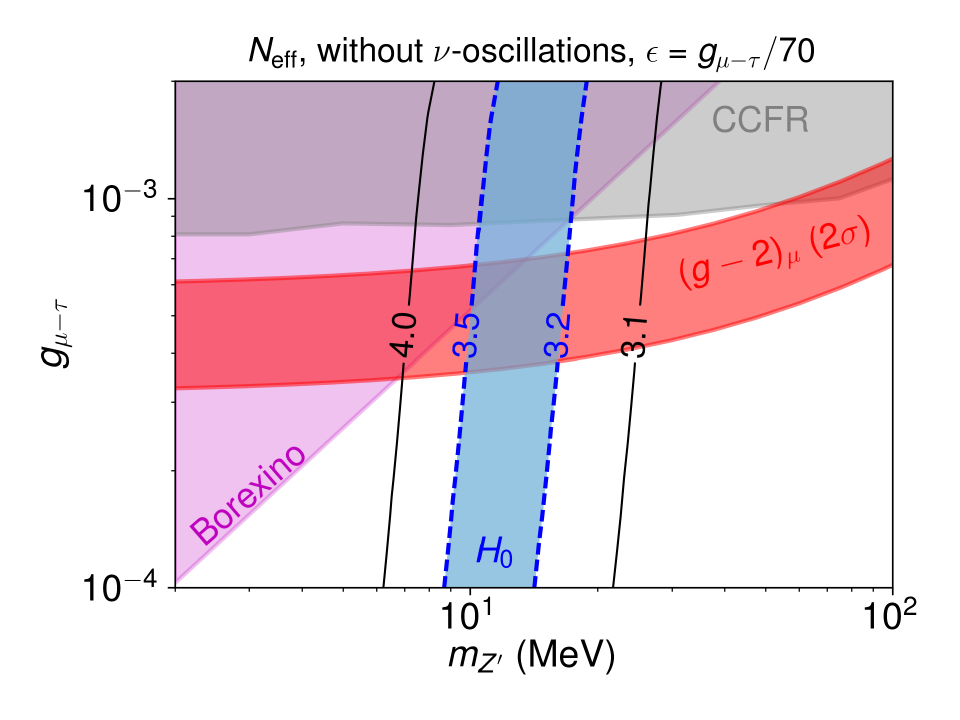} \\
\includegraphics[width=0.45\textwidth]{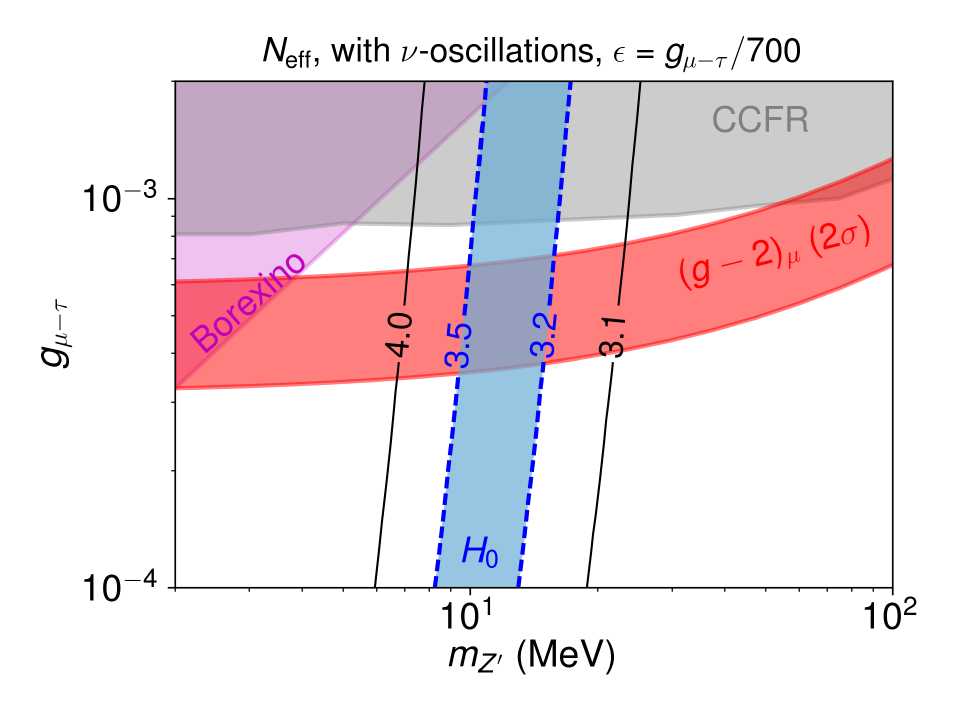} &\includegraphics[width=0.45\textwidth]{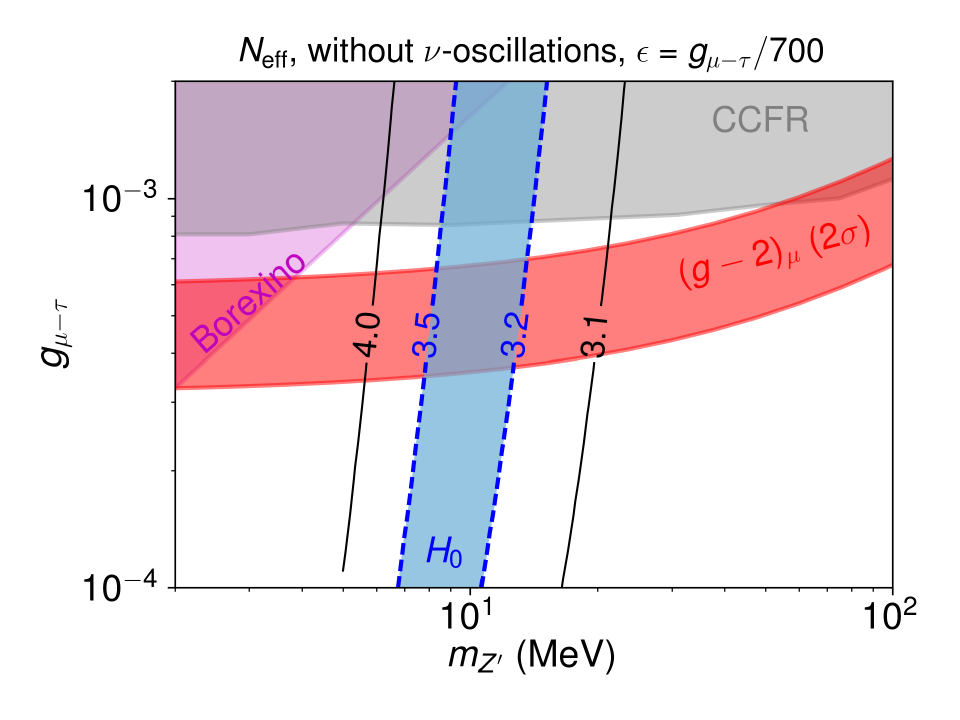}\\
\includegraphics[width=0.45\textwidth]{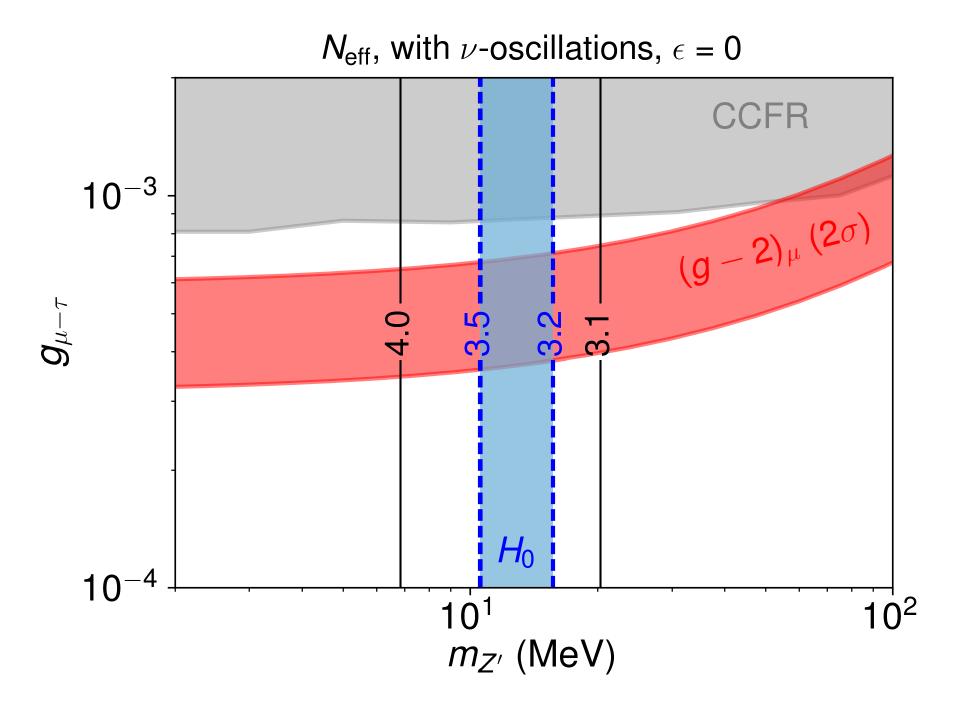} &\includegraphics[width=0.45\textwidth]{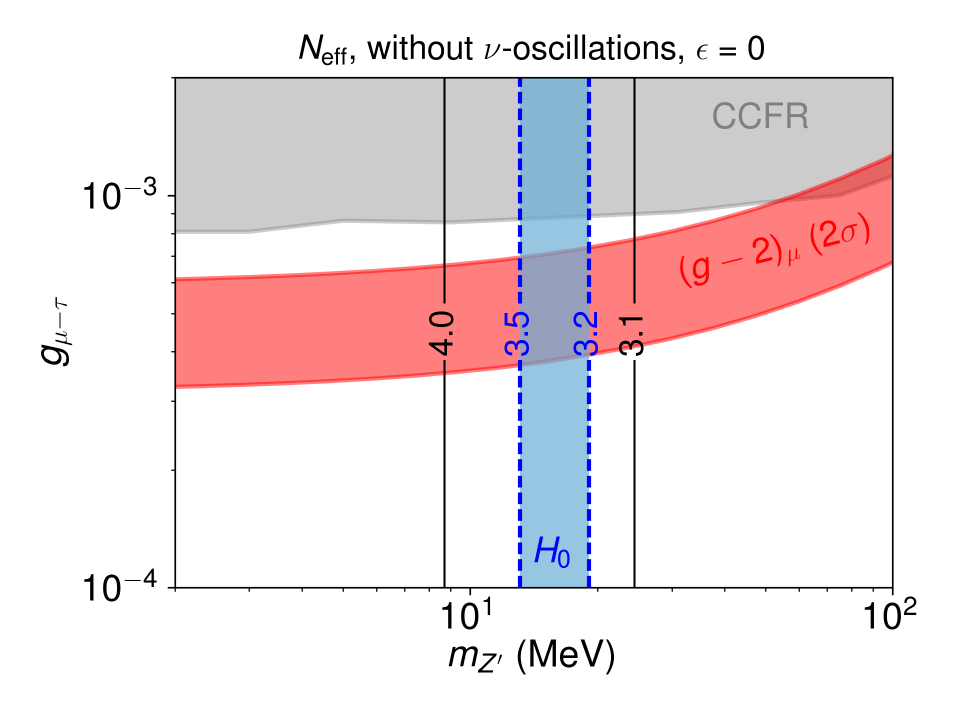}
\end{tabular}
\caption{Contours of constant $N_\text{eff}$ as a function of the $\zp$ mass and gauge coupling for three different values of the kinetic mixing parameter, $\epsilon$. In the left panels we assume that neutrino oscillations equilibrate the $\nu_e$ component such that $T_{\nu_e} = T_{\nu_\mu}$, whereas neutrino oscillations are neglected in the right frames. We also show are the regions of parameter space favored by measurements of the muon's anomalous magnetic moment, $(g-2)_\mu$, as well as those excluded by measurements of di-muon production in neutrino nucleus scattering (CCFR)~\cite{Mishra:1991bv,Altmannshofer:2014pba}, and solar neutrino scattering  (Borexino)~\cite{Harnik:2012ni,Laha:2013xua,Kamada:2015era} (see Sec.~\ref{sec:constraints}).
This parameter space is also constrained rare kaon decays, but the bound is weaker than the CCFR region above \cite{Ibe:2016dir}. 
Intriguingly, the full parameter space that simultaneously ameliorates both the $H_0$ and $(g-2)_\mu$ anomalies 
can also be tested with fixed target experiments \cite{Gninenko:2018tlp,Kahn:2018cqs,Berlin:2018bsc,Chen:2018vkr}, rare kaon decays \cite{Ibe:2016dir,diego}, and at Belle-II~\cite{Araki:2017wyg,Banerjee:2018mnw}.
 }
\label{fig:results_large_g}
\end{figure}

\begin{figure}[t]
\centering
\begin{tabular}{cc}
\includegraphics[width=0.49\textwidth]{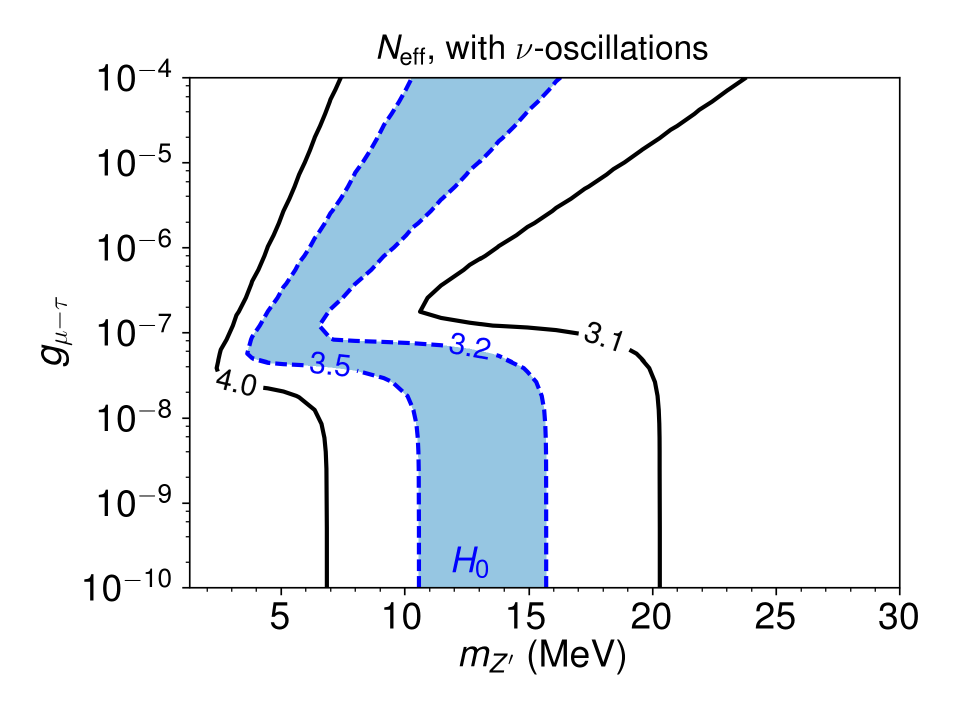} & \includegraphics[width=0.49\textwidth]{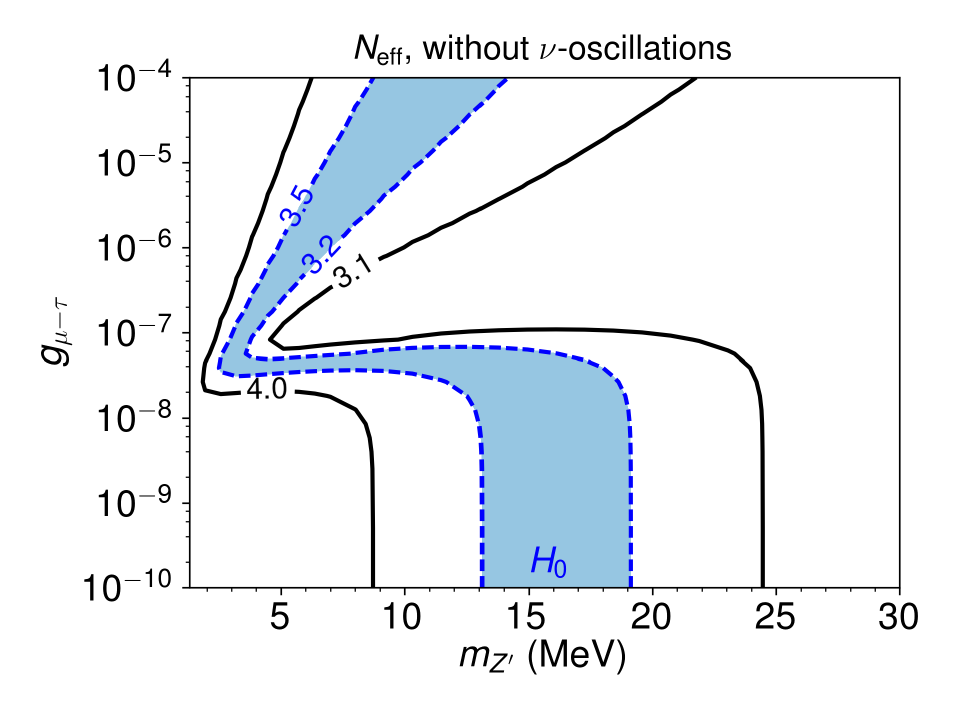}
\end{tabular}
\caption{As in Fig.~\ref{fig:results_large_g}, but for smaller values of $g_{\mu-\tau}$ and for the case of $\epsilon = g_{\mu-\tau}/70$.}
\label{fig:results_Neff}
\end{figure}

Using this set of equations, we can describe the process of neutrino decoupling, including the energy injection into the SM plasma from $\zp$ decays and any exchange of energy between neutrinos and the electromagnetic sector that might result from $U(1)_{L_{\mu}-L_{\tau}}$ interactions. We solve this set of equations starting from an initial condition of $T_\gamma = T_\nu = 20\,\text{MeV}$, for which the SM weak interactions are very efficient ($\Gamma/H \sim 300$). In Fig.~\ref{fig:Tplots} we plot the evolution of the photon-to-neutrino temperature ratio for several choices of the $\zp$ mass and coupling. Note that when solving these equations we explicitly check that the continuity equation, $\dot{\rho} = -3H(\rho +p)$, is fulfilled to a relative accuracy of at least $10^{-5}$. In Fig.~\ref{fig:results_large_g} we show contours for the value of $N_{\rm eff}$ predicted across a range of parameter space in this model, focusing on the range of masses and couplings that are potentially relevant for the $(g-2)_\mu$ anomaly. We show these results for three choices of the kinetic mixing parameter, and both including or ignoring the impact of neutrino oscillations. Also shown are the constraints on this parameter space derived from measurements of di-muon production in neutrino nucleus scattering (CCFR)~\cite{Mishra:1991bv,Altmannshofer:2014pba}, the solar neutrino scattering rate (Borexino)~\cite{Harnik:2012ni,Laha:2013xua,Kamada:2015era} (see Sec.~\ref{sec:constraints}). For $m_{\zp} \sim 10-20$ MeV, the predicted value of $N_{\rm eff}$ falls within the range capable of relaxing the tension between late and early-time determinations of the Hubble constant. Furthermore, for $g_{\mu-\tau} \simeq (3-8) \times 10^{-4}$, such a particle can also account for the observed value of $(g-2)_{\mu}$.  In Fig.~\ref{fig:results_Neff}, we extend these results to smaller values of $g_{\mu-\tau}$. Although this parameter space cannot address the measured magnetic moment of the muon, the Hubble tension can be relaxed for a $\zp$ with a wide range of couplings.

Before moving on, we note that the results described in this section do not agree with those presented in Ref.~\cite{Kamada:2015era}. This is the case for two primary reasons. First, the authors of Ref.~\cite{Kamada:2015era} did not consider kinetic mixing between the photon and the $\zp$ and thus neglect potentially important process of the form $e^+ e^- \leftrightarrow Z'$. Second, they adopted 1.5 MeV for the decoupling temperature of the $\nu_{\mu,\tau}$ neutrinos, instead of the correct value of $T_{\nu_\mu}^\text{dec} \simeq 3.2\,\text{MeV}$ (or $T\simeq 2.4\,\text{MeV}$ after taking into account oscillations)~\cite{Escudero:2018mvt,Hannestad:2001iy,Dolgov:2002ab,Dolgov:2002wy}. If we set $\epsilon=0$ and $T^{\rm dec}_{\nu_{\mu},\nu_{\tau} } = 1.5$ MeV, we are able to reproduce the results presented in Ref.~\cite{Kamada:2015era}. In a follow up study, Ref.~\cite{Kamada:2018zxi} considered the interactions $e^+ e^- \leftrightarrow Z' $ in the particular case in which $g_{\mu-\tau} = 5 \times 10^{-4}$ and used entropy conservation to track the temperature evolution of the different sectors. Our results are similar to those presented in Ref.~\cite{Kamada:2018zxi} for $g_{\mu-\tau} = 5 \times 10^{-4}$. Ref.~\cite{Kamada:2018zxi}, however, did not include SM neutrino-electron interactions and assumed $T_{\nu_\mu}^\text{dec} = 1.5\,\text{MeV}$, whereas the present work includes these interactions and calculates the value of $T_{\nu}^{\rm dec}$.

\subsection{Freeze-In (Late Equilibration) Regime} 
\label{sec:freezein}

If the value of $g_{\mu-\tau}$ is very small, the $\zp$ population will not reach equilibrium with the SM thermal bath in the very early universe. The condition for these particles to reach equilibrium with muons and taus at early times can be written as $n_\mu \langle \sigma v \rangle_{\mu \mu \to \gamma \zp} > H(m_\mu)$, corresponding to the following requirement on the gauge coupling:
\be
\label{eq:mu-thermalization}
\gmt \gsim   \left( \frac{  1.66 \sqrt{g_*} }{\alpha}  \frac{m_\mu}{    m_{ \rm Pl}    }  \right)^{1/2} \simeq 4 \times 10^{-9} ,
\ee
where we have approximated $ \langle \sigma v \rangle_{\mu \mu \to \gamma \zp} \sim \alpha g_{\mu-\tau}^2/ 2m_\mu^2$ in 
the nonrelativistic limit near $T \sim m_\mu$, where the production rate is maximized relative to that of Hubble expansion. As long as $m_\zp \ll m_\mu$, the condition in Eq.~(\ref{eq:mu-thermalization}) is independent of $m_\zp$.

But even if $g_{\mu-\tau}$ is too small for the $\zp$ population to be thermalized through $\mu^+ \mu^- \to Z' \gamma$ annihilations, a significant abundance of $Z'$'s can be produced through inverse decays. In fact, inverse decays may ultimately be able to generate an equilibrium $\zp$ abundance. More specifically, in the case of a light $\zp$ (for which inverse decays are active when $T_\nu \gg m_{Z'}$), equilibrium between the $\zp$ and neutrino populations will be achieved for $g_{\mu-\tau} \gsim 1.3 \times 10^{-10} (m_{\zp}/{\rm MeV})^{1/2}$.

After neutrino decoupling, neutrino oscillations are efficient enough to approximately equilibrate the neutrino distributions~\cite{Dolgov:2002ab,Hannestad:2001iy,Dolgov:2002wy}, and thus even if the $Z'$ only interacts with muon and tau neutrinos the rapid oscillation will render the $\nu_e$ distribution very similar to that of $\nu_\mu$ and $\nu_\tau$. Taking this into account, and imposing energy and number density conservation, we can calculate the neutrino and $Z'$ distributions after equilibration. These equations explicitly read
\be 
3 \rho_{\nu}(T_\nu,0)  &=&  3 \rho_{\nu}(T_{\rm eq},\mu_{\rm eq})   + \rho_{\zp}(T_{\rm eq},2\mu_{\rm eq})  , \\
3 n_{\nu}(T_\nu,0)  &=&  3 n_{\nu}(T_{\rm eq},\mu_{\rm eq})   + 2 n_{\zp}(T_{\rm eq},2\mu_{\rm eq})�\, ,
\ee 
where the equilibrium condition requires $\mu_{Z'} = 2 \mu_\nu$. Solving these equations simultaneously yields the following result:
\be \label{eq:thermal_T_mu}
\frac{T_{\rm eq}}{T_{\nu}} = 1.2076 \,, &\qquad& \frac{\mu_{\rm eq}}{T_{\nu}} = 1.1664 \, .
\ee
Thus the neutrino temperature increases at the same time that a chemical potential is developed for the neutrino and $Z'$ populations. Once the $Z'$ population is in equilibrium with the neutrinos, it will maintain its distribution due to the active decays and inverse decays until the temperature drops to $T_\nu \sim m_{Z'}/3$, at which time they will decay out of equilibrium and thereby generate a net contribution to $N_\text{eff}$. Since this decay process is occurring out of equilibrium, we cannot use Fermi-Dirac or Bose-Einstein distributions but instead have to solve the full Boltzmann equation for the distribution functions. The relevant Boltzmann equations read as follows~\cite{Kawasaki:1992kg,Dolgov:1998st}:
 \begin{eqnarray} 
\frac{\partial f_{Z'} }{\partial t}  - H p_{Z'}  \frac{\partial f_{Z'} }{\partial p_{Z'}}   &=& - \frac{m_{Z'} \Gamma_{Z'}}{E_{Z'} p_{Z'}}  \int_{\frac{E_{Z'} -p_{Z'}}{2}}^{\frac{E_{Z'} + p_{Z'}}{2}} d E_\nu F_{\rm dec}\left( E_{Z'}, E_\nu, E_{Z'}-E_\nu \right) \, , \label{eq:ftosolve}\\
 \frac{\partial f_{\nu} }{\partial t}  - H p_{\nu}  \frac{\partial f_{\nu} }{\partial p_{\nu}}  &=& \frac{m_{Z'} \Gamma_{Z'}}{E_{\nu} p_{\nu}} \int_{|(m^2_{Z'}/4p_{\nu})-p_{\nu}|}^{\infty} \frac{d p_{Z'}  p_{Z'}}{E_{Z'}} F_{\rm dec}\left( E_{Z'}, E_\nu, E_{Z'}-E_\nu \right) \, , \label{eq:ftosolve_fnu}
 \end{eqnarray}
 where 
 \begin{eqnarray}\label{eq:boltzdecay_Fdecay}
F_{\rm dec}\left( E_{Z'}, E_{\nu_1}, E_{\nu_2} \right) =  f_{Z'}(E_{Z'})  \left[1-f_\nu(E_{\nu_1})\right]  \left[1-f_\nu(E_{\nu_2})\right] -f_\nu(E_{\nu_1}) f_\nu(E_{\nu_2}) \left[ 1+f_{Z'}(E_{Z'})\right].
 \end{eqnarray}
 By using the continuity equation we find that the photon temperature simply evolves as 
\be
  \label{eq:boltzdecay_Fdecay}
  \frac{dT_{\gamma}}{dt}  =- \biggl[  4 H \rho_{\gamma} + 3 H ( \rho_{e} + p_{e} )     \biggr]   \left(    \frac{\partial \rho_{\gamma}}{\partial T_\gamma} + \frac{\partial \rho_e}{\partial T_\gamma}       \right)^{-1}   .
\ee
In order to solve Eqns.~(\ref{eq:ftosolve}) and~(\ref{eq:ftosolve_fnu}), we bin the $Z'$ and $\nu$ distribution functions in comoving momentum, $y = a \,p/\text{MeV}$, from $ y_\text{min} = 0.01$ to $y_\text{max}=20$ in 100 bins (this choice has been previously shown to produce accurate results~\cite{Dolgov:1997mb,Dolgov:1998st}). We use this set of equations for both the equilibration and freeze-in cases, since out-of-equilibrium decays result in each. For the freeze-in case, we start the integration at a temperature of $T_\nu = 100\times m_{Z'}$ with the initial condition $f_{Z'} = 0$ and $f_\nu = f^{\rm FD}(T_\nu,\mu_\nu = 0)$, whereas in the equilibrium case we start the integration at a temperature of $T_\nu = 10\times m_{Z'}$ and with the initial condition $f_{Z'} = f^{\rm BE}(T_{\rm eq},2 \, \mu_\nu)$ and $f_\nu = f^{\rm FD}(T_{\rm eq}, \mu_\nu)$. 

\begin{figure}[t]
\centering
\begin{tabular}{c}
\vspace{-0.6cm}
\includegraphics[width=0.6\textwidth]{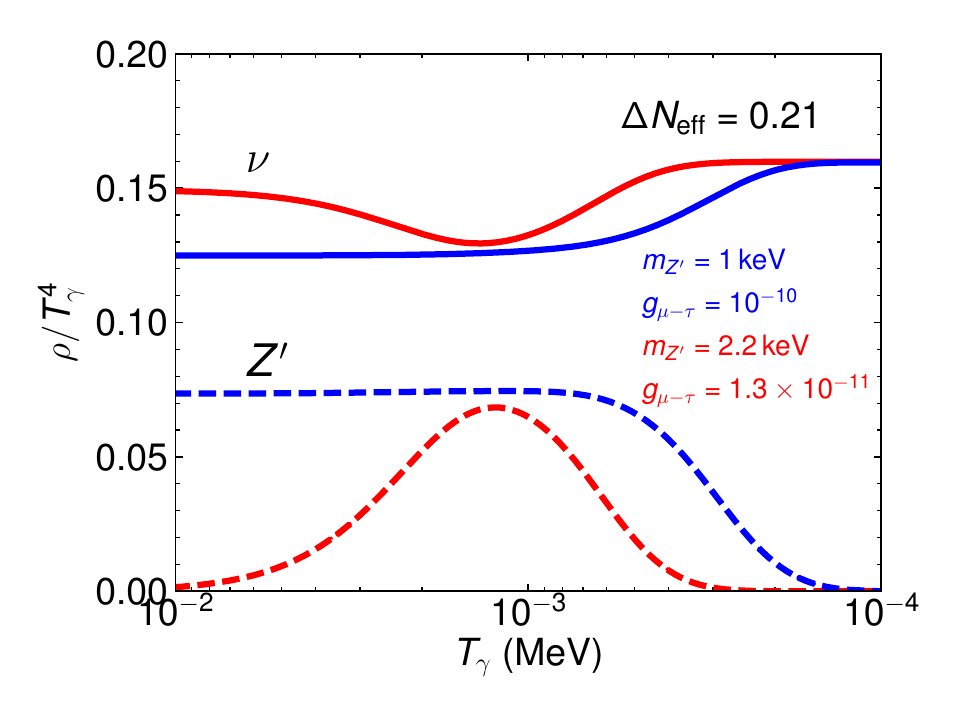}
\end{tabular}
\caption{The evolution of the neutrino (solid) and $Z'$ (dashed) energy densities in a regime in which only decays and inverse decays are effective and occur after neutrino decoupling. The red lines correspond to a situation in which the $Z'$ thermalizes at a temperature similar to its mass $T_\nu \sim m_{Z'}$, prior to which its energy density was negligible. In blue we show a case in which the $Z'$ population equilibrates with the neutrinos while still relativistic.}\label{fig:evZpfreezeIN}
\end{figure}

In Fig.~\ref{fig:evZpfreezeIN} we show the evolution of the neutrino and $\zp$ energy densities in two representative cases. In the case of $g_{\mu-\tau} = 10^{-10}$ and $m_{Z'} = 1\,\text{keV}$, the $Z'$ population thermalizes with the neutrinos while relativistic and only decays after $T_\nu \sim m_{Z'}/3$, rendering $\Delta N_\text{eff} \simeq 0.21$. By contrast, for the case of $g_{\mu-\tau} = 1.3\times10^{-11}$ and $m_{Z'} = 2.2\,\text{keV}$, the $Z'$ population reaches thermal equilibrium with the neutrinos at $T_\nu \sim m_{Z'}$. We see in this case that the neutrino energy density is reduced as the $Z'$ population grows, and then increases as the $\zp$ population decays, to yield $\Delta N_\text{eff} \simeq 0.21$\footnote{If the effect of neutrino oscillations is neglected in this computation, we find $\Delta N_\text{eff} \simeq 0.18$ instead.}. In Appendix~\ref{app:App_Freeze-In} we display the time evolution of $f_{Z'}$ and the final $f_\nu|_{T_\gamma \ll m_{Z'}}$ for some representative values of $g_{\mu-\tau}$ and $m_{Z'}$.

In exploring the freeze-in parameter space within this model, we find the following behavior:
\begin{align}
\Gamma_{Z'} \geq H(T_\nu = m_{Z'}) \Longrightarrow \,&\Delta N_{\rm eff} \simeq 0.21�\nonumber \\
\Gamma_{Z'} = H(T_\nu = m_{Z'})/5^2 \Longrightarrow \,&\Delta N_{\rm eff} \simeq 0.1�\\
\Gamma_{Z'} = H(T_\nu = m_{Z'})/17^2 \Longrightarrow \,&\Delta N_{\rm eff} \simeq 0.03�\nonumber
\end{align}

We would like to emphasize that throughout the parameter space in which the $\zp$ population reaches thermal equilibrium with the neutrinos, this model yields a prediction of $\Delta N_\text{eff} \simeq 0.21$. This limit is realized across a wide range of the $m_{\zp}-g_{\mu-\tau}$ parameter space (see Fig.~\ref{fig:money}), making $\Delta N_\text{eff} \simeq 0.21$ a common feature within the context of this model.

We appreciate that this result will strike many readers as counterintuitive. To understand why this $\Delta N_{\rm eff} \simeq 0.21$ plateau arises within the parameter space of this model, consider the following summary. If $g_{\mu-\tau} \lsim 4 \times 10^{-9}$, the $\zp$ population will not thermalize with muons and taus at early times (see Eq.~\ref{eq:mu-thermalization}). But so long as $g_{\mu-\tau} \gsim 1.3 \times 10^{-10} \, (m_{\zp}/{\rm MeV})^{1/2}$, a light $\zp$ will ultimately obtain an equilibrium abundance through inverse decays. Thus for the a sizable range of couplings that fall in between these two limits, the $\zp$ population will not reach equilibrium with muons or taus at early times, but will equilibrate with neutrinos, ultimately leading to $\Delta N_{\rm eff} \simeq 0.21$ through their decays. Note that  $\Delta N_{\rm eff} \simeq 0.21$ will only result from $Z'$'s with $m_{Z'} \gtrsim 1\,\text{eV}$, since for lighter gauge bosons the $Z'$ population would not have decayed prior to $T_{\rm CMB} \simeq 0.26\,\text{eV}$.




\section{Constraints}
\label{sec:constraints}

In this section, we consider a number of constraints on this model, including those that apply directly to the gauge coupling, and which depend on the degree of kinetic mixing between the $\zp$ and the photon.\footnote{We have explicitly checked that given the small branching fraction of the $Z'$ to electrons $ {\rm Br}_{Z'\to e^+e^-} \simeq 2\times10^{-5}$~\eqref{eq:branching}, the energy injection from the $Z'$ population decay during the CMB or BBN~\cite{Fradette:2014sza} is too small to result in meaningful constraints on the model.} See \textit{e.g.}~\cite{Bauer:2018onh,Ilten:2018crw} for an exhaustive set of constraints for MeV-GeV $Z'$'s.
\subsection{Neutrino Tridents}
\label{sec:tridents}

Light $L_\mu - L_\tau$ gauge bosons are constrained by the results of the Columbia-Chicago-Fermilab-Rochester (CCFR) Neutrino Experiment, which are consistent with the SM predictions for $\nu N \to \nu N \mu^+ \mu^-$ trident production~\cite{Mishra:1991bv,Altmannshofer:2014pba}. 
In the presence of a light $\zp$, this process receives additional corrections from the three body $\nu N \to \nu N \zp$ process followed by a prompt $\zp \to \mu^+\mu^-$ decay. 
Consistency with the CCFR trident measurement requires $\gmt \lsim 10^{-3}$ for $m_\zp \ll \MeV$, as shown in Figs.~\ref{fig:money} and~\ref{fig:results_large_g}.

\subsection{Supernova 1987A}
\label{sec:SN}

A light $\zp$ can be produced via neutrino inverse decays $\bar \nu \nu \to \zp$ within the dense core of supernovae. If such particles are long-lived, they may efficiently remove energy from the core and thus modify the observed $\sim$\,10 second duration of the neutrino burst from Supernova 1987A~\cite{Kolb:1987qy}. 
 In Figs.~\ref{fig:money} we show the regions of parameter space in which the $\zp$ luminosity introduces an order one correction to the total energy loss ($\sim$\,$10^{53}$ erg) in the first $\sim$\,10 seconds of a supernova explosion, assuming a $\sim$\,10 km radius and a $\sim$\,30 MeV core temperature. The details of this estimate are described in Appendix~\ref{app:SNbounds}.

\subsection{Stellar Cooling}
\label{subsec:stellarcooling}

In this model, the $\zp$ can be produced in 
stellar plasmas and induce anomalous cooling by
carrying energy away from stars.
Since this process yields an invisibly decaying particle, the 
stellar bounds on dark photons from Refs.~\cite{An:2013yfc,Hardy:2016kme} can straightforwardly be adapted to constrain the $L_{\mu}-L_{\tau}$ model. This constraint is shown in the lower frame of Fig.~\ref{fig:money}. We also note that bounds from white dwarfs cooling apply to our model~\cite{Bauer:2018onh,Dreiner:2013tja}, and that they are of comparable strength to those derived from Borexino for $Z'$'s in the MeV mass range~\cite{Kamada:2018zxi}. However, such constraints have been derived assuming that the $Z'$ is more massive than the typical temperature of a white dwarf ($T \sim 5\,\text{keV}$), and therefore it is not trivial to extrapolate them for $m_{Z'} \lesssim 10\,{\rm keV}$, and hence are not shown in the figures.

\subsection{Solar Neutrino Scattering}
\label{subsec:SolarNeutrinos}

Nonzero kinetic mixing introduces $\zp$ mediated interactions between $\nu_{\mu,\tau}$ and 
charged fermions, which can distort the observed solar neutrino scattering rate at Borexino~\cite{Harnik:2012ni,Laha:2013xua,Kamada:2015era}. These constraints are shown in Figs.~\ref{fig:money} and~\ref{fig:results_large_g} (in those frames corresponding to $\epsilon >0)$. 




\section{Summary and Conclusions}
\label{sec:conclusion}

The reported tension between early and late time determinations of the Hubble constant has motivated us to consider models that include a new light gauge boson. In particular, in light of the stringent constraints on gauge bosons with couplings to first generation fermions, we consider a model with a broken $U(1)_{L_\mu - L_\tau}$ gauge symmetry, corresponding to a massive gauge boson with tree-level couplings to muons, taus, and their corresponding neutrino species. We have studied the impact of such a particle on the evolution of the early universe, solving the full set of Boltzmann equations and determining the contribution to the energy density in radiation, parameterized in terms of $\Delta N_{\rm eff}$.

We have identified two distinct regions of parameter space in which the tension related to the Hubble constant can be substantially relaxed (corresponding to $\Delta N_{\rm eff} \sim 0.2-0.5$):

\begin{itemize}

\item{For a gauge boson with $g_{\mu-\tau} \gsim 4\times 10^{-9}$, this particle will reach equilibrium with the Standard Model bath in the early universe. For $m_{\zp} \sim$\,$10-20$ MeV, the decays of these particles will heat the neutrino population and delay the process of neutrino decoupling, resulting in $\Delta N_{\rm eff} \sim 0.2-0.5$. For a coupling of $g_{\mu-\tau} \simeq (3-8) \times 10^{-4}$, such a $\zp$ could also account for the measured value of the muon's anomalous magnetic moment. We also note that the parameter space which simultaneously ameliorates both the $H_0$ and $(g-2)_\mu$ anomalies can also be tested with existing and future accelerator searches for muon specific forces with muon beams \cite{Gninenko:2018tlp,Kahn:2018cqs,Berlin:2018bsc,Chen:2018vkr}, measurements of rare kaon decays \cite{Ibe:2016dir,diego}, and at Belle-II~\cite{Araki:2017wyg,Banerjee:2018mnw}.}

\item{For a very light ($m_{Z'} \gtrsim 1\,\text{eV}$) gauge boson with a very small coupling ($g_{\mu-\tau} \lsim 4\times 10^{-9}$), the $\zp$ population does not reach equilibrium with muons or taus at early times. Instead, a significant abundance of such particles can be produced through $\bar \nu \nu \to \zp$ inverse decays. In particular, for $g_{\mu-\tau} \gsim 1.3 \times 10^{-10} (m_{\zp}/{\rm MeV})^{1/2}$, inverse decays will ultimately produce a $\zp$ population that reaches equilibrium with neutrinos. When this $\zp$ population decays, it produces an energy density of neutrinos corresponding to $\Delta N_{\rm eff} \simeq 0.21$. This result is found across a wide range of parameter space within this model.}

\end{itemize}

These results are summarized in Fig.~\ref{fig:money}, which includes contours of constant $\Delta N_\text{eff}$, as well as the regions favored by measurements of the muon's anomalous magnetic moment, $(g-2)_\mu$. Among other motivations, this model is particularly interesting in light of the large regions of its parameter space that can relax the reported Hubble tension ($\Delta N_{\rm eff} \sim 0.2-0.5$) and that are within the reach of next-generation CMB measurements ($\Delta N_{\rm eff} \gsim 0.03$~\cite{Abazajian:2016yjj}).

\section*{Acknowledgments}
We thank Asher Berlin, Sam McDermott for helpful conversations. 
Fermilab is operated by Fermi Research Alliance, LLC, under Contract No. DE- AC02-07CH11359 with the US Department of Energy. This project has received funding/support from the European Union's Horizon 2020 research and innovation programme under the Marie Skłodowska-Curie grant agreements Elusives ITN No. 674896 and InvisiblesPlus RISE No. 690575. The work of MP was supported by the Spanish Agencia Estatal de Investigaci\'{o}n through the grants FPA2015-65929-P (MINECO/FEDER, UE), IFT Centro de Excelencia Severo Ochoa SEV-2016-0597, and Red Consolider MultiDark FPA2017-90566-REDC. ME is supported by the European Research Council under the European Union's Horizon 2020 program (ERC Grant Agreement No 648680 DARKHORIZONS). ME and MP acknowledge the Fermilab Theory and Astro groups for their hospitality when this project was initiated. 
\medskip


\medskip

\bibliography{biblio}

\newpage
\appendix  


\renewcommand{\theequation}{A.\arabic{equation}}
\setcounter{equation}{0}
\section{Boltzmann Equation for $\zp$ Production}\label{app:Boltzmann_Zp}
\label{app:A}

The number density evolution of the $Z'$ in the presence of the $\nu(p_1) + \nu(p_2) \longleftrightarrow Z^\prime (p_3)$ process is  
\be
\frac{ dn_{\zpr}}{dt}  +  3 H n_{\zp} = C[f_\zpr],
\ee
where $n_{\zp}$ is the $\zp$ number density, $H$ is the Hubble rate, and the collision term satisfies 
\be
C[f_\zpr] =  \int  \frac{d^3 \vec p_1}{(2\pi)^3 2 E_1} \frac{d^3 \vec p_2}{(2\pi)^3 2 E_2} \frac{d^3 \vec p_\zp}{(2\pi)^3 2 E_\zp} |{\cal A}_{\zpr \to \nu\bar \nu}|^2\left[ f_\nu(\vec p_1) f_\nu(\vec p_2)   -     
 f_\zpr(\vec p_\zp)            \right] (2\pi)^4 \delta^4(p_1+p_2 - p_\zp) ,~~~~~
\ee
where $|{\cal A}_{i \to f}|^2$ is the squared, spin-averaged matrix element for the process in the subscript.  By detailed balance, we can simplify the inverse process using 
\be
 f^{\rm (eq)}_\nu(\vec p_1) f^{\rm (eq)}_\nu(\vec p_2)   =  f^{\rm (eq)}_\zp(\vec p_\zp)~,
\ee
so the collision terms can be simplified
\be
 \int  \frac{d^3 \vec p_1}{(2\pi)^3 2 E_1} \frac{d^3 \vec p_2}{(2\pi)^3 2 E_2} \frac{d^3 \vec p_\zp}{(2\pi)^3 2 E_\zp}     |{\cal A}_{\zpr \to \nu\bar \nu}|^2    \left[  f^{\rm (eq)}_\zp(\vec p_\zp)  -     
 f_\zp(\vec p_\zp)             \right] (2\pi)^4 \delta^4(p_1+p_2 - p_\zp).~~~~
\ee
Using the definition of the {\it energy dependent} width
\be
\Gamma_{\zpr}(E_\zp) = \frac{1}{2E_\zp} \int  \frac{d^3 \vec p_1}{(2\pi)^3 2 E_1} \frac{d^3 \vec p_2}{(2\pi)^3 2 E_2} |{\cal A}_{\zpr \to \nu \bar \nu}|^2 
        (2\pi)^4 \delta^4(p_1+p_2 - p_\zp) ,~~~
\ee
where the prefactor is $1/E_\zp$ instead of $1/m_{\zp}$ (as in the rest frame expression), the Boltzmann equation becomes
\be \label{eq:boltzt}
s \frac{dY_{\zpr}}{dt}  =
 \int   \frac{d^3 \vec p_\zp}{(2\pi)^3 2 E_\zp}  2 E_\zp  \Gamma_{\zpr}(E_\zp)    \left[  f^{\rm (eq)}_3(\vec p_\zp)  -     
 f_3(\vec p_\zp)             \right] = \langle     \Gamma_{\zpr}     \rangle   \left[  n_{\zpr}^{\rm (eq)}(t)  -    n_{\zpr}(t) \right],
 \ee
where we have defined the comoving dimensionless yield $Y_{\zp} \equiv n_{\zp} / s $ and   
\be
\langle  \Gamma_\zpr\rangle \equiv \frac{\int  d^3 \vec p_\zp  \,     \Gamma(E_\zp)  e^{-E_\zp/T}    }{
\int   d^3 \vec p_\zp   \,  e^{-E_\zp/T}} 
=  \frac{ \Gamma_{\zpr}}{ m_\zpr T K_2\left( \frac{m_\zpr}{T}\right)}  \int_{m_\zpr}^\infty dE_\zp \sqrt{E_\zp^2 - m_\zpr^2} e^{-E_\zp/T}  
=   \Gamma_{\zpr}  \frac{ K_1\left( \frac{m_\zpr}{T}\right)  }{  K_2\left( \frac{m_\zpr}{T}\right)}  ~,~~~~
\ee
which is valid in the limit of Boltzmann statistics. 
Note that Eq.~(\ref{eq:boltzt}) has the familiar freeze out form, except that there is a linear dependence on the phase space distributions (as opposed to quadratic) and the thermally averaged cross section for annihilation is 
replaced with a thermally averaged decay rate. 
Defining $z \equiv m_{\zp} / T$, and $Y_{Z'} = n_{Z'}/s$, where $s$ is the total entropy density, we have
 \be
 H(z) = \frac{ 1.66 \sqrt{g_*}  }{m_{Pl}}   \left( \frac{m_{\zp}}{z} \right)^2~~, ~~~ s(z) = \frac{2 \pi^2}{45} g_{*S} \left( \frac{m_{\zp}}{z} \right)^3~~,~~  
 Y_{\zpr}^{\rm (eq)}(z)  = \frac{g_{\zp}}{2 \pi^2 s(z)}       \int_{m_{\zp} }^\infty   dy   \frac{ y        \sqrt{y^2 - m^2_{\zp}  }    }{ e^{y z /m_{\zp}}    -1 },~~~~~~
 \ee
and  we can rewrite the full Boltzmann equation as 
\be \label{eq:full-boltz}
 \frac{dY_{\zpr}}{dz}  =  \frac{   \Gamma_{\zpr}  }{ H(z)  z  } \frac{ K_1(z) }{  K_2(z) }   \left[  Y_{\zpr}^{\rm (eq)}(x)  -    Y_{\zpr}(x) \right],
 \ee
which is equivalent to Eq.~(\ref{eq:zp-boltz}) expressed in terms of the yield, $Y_{Z'}$. Note that this equation accounts for both production and decay processes. 

\section{Supernova 1987A Bounds}\label{app:SNbounds}
\renewcommand{\theequation}{B.\arabic{equation}}
\setcounter{equation}{0}
\label{sec:appB}

The energy-density loss rate per unit volume from $\nu(p_1) + \nu(p_2) \to Z^\prime (p_{\zp})$ production
in an isotropic, homogeneous supernova is  
\be
 \frac{d\rho_{\zpr}}{dt} =  \int  \frac{d^3 \vec p_1}{(2\pi)^3 2 E_1} \frac{d^3 \vec p_2}{(2\pi)^3 2 E_2} \frac{d^3 \vec p_{\zp} }{(2\pi)^3 2 E_{\zp}}   E_{\zp} \langle |{\cal A}_{\zpr \to \nu\bar \nu}|^2 \rangle     f^{\rm (eq)}_\nu(\vec p_1) f_\nu^{\rm (eq)}(\vec p_2)  (2\pi)^4 \delta^4(p_1+p_2 - p_{\zp}).~~~~~~
\ee
Since the neutrinos are in equilibrium in the core, but the $\zp$ population is not, we neglect the reverse process. By detailed balance, we can substitute $f^{\rm (eq) }_\nu(\vec p_1)   f^{\rm (eq)}_\nu(\vec p_2)   =     
 f^{\rm (eq)}_{\zp}(\vec p_3) $, which yields
 \be
 \frac{d\rho_{\zpr}}{dt} =  \int  \frac{d^3 \vec p_1}{(2\pi)^3 2 E_1} \frac{d^3 \vec p_2}{(2\pi)^3 2 E_2} \frac{d^3 \vec p_{\zp} }{(2\pi)^3 2 E_{\zp}}   E_{\zp} \langle |{\cal A}_{\zpr \to \nu\bar \nu}|^2 \rangle  f^{\rm (eq)}_{\zp}(\vec p_{\zp})   (2\pi)^4 \delta^4(p_1+p_2 - p_{\zp}) .~~~~~~
\ee
Since we are mainly interested in very small $\zp$ couplings in the non-equilibrium regime, we can neglect the reverse process which depends on $f_{\zp}$, so using the 
definition of the rest frame width
 \be
\Gamma_{\zp}= \frac{1}{2 m_{\zp}} \int  \frac{d^3 \vec p_1}{(2\pi)^3 2 E_1} \frac{d^3 \vec p_2}{(2\pi)^3 2 E_2}  \langle |{\cal A}_{\zpr \to \nu\bar \nu}|^2 \rangle    (2\pi)^4 \delta^4(p_1+p_2 - p_{\zp}) ,~~~
\ee
the total production rate can be written
\be
 \frac{d\rho_{\zpr}}{dt} =  m_{\zp} \Gamma_{\zp} \int   \frac{d^3 \vec p_{\zp}}{(2\pi)^3}        
 f^{\rm (eq)}_{\zp}(\vec p_{\zp})         ,~~~
\ee
where the integral corresponds to the equilibrium number density of $\zp$s inside the supernova. To calculate the total energy carried away by the  $\zp$ population, we need to include the probability that a given $\zp$ will survive out to a distance of $R \sim 10$ km at $T \sim 30$ MeV, so we have
\be
 \frac{d\rho_{\zpr}}{dt} =  m_{\zp} \Gamma_{\zp} \int   \frac{d^3 \vec p_{\zp}}{(2\pi)^3}        
 f^{\rm (eq)}_{\zp}(\vec p_{\zp})    \,  e^{-\Gamma_{\zp} R/\gamma}  = 
\frac{ 3 \, m_{\zp} \Gamma_{\zp} }{2\pi^2}  \int_0^\infty   d  p_{\zp} p_{\zp}^2          
 e^{-E_{\zp}/T}  \,  e^{-\Gamma_{\zp} R/\gamma} 
\ee
where $\gamma = E_{\zp}/m_{\zp}$ is the boost factor and the factor of 3 accounts for the number of polarization states in the convention $f^{\rm (eq)} = 3\, e^{-E_{\zp}/T}$ since we performed a spin average in the definition of the rest frame width.
To ensure that we don't modify the observation of Supernova 1987A by an order one amount, we demand that the total energy loss in a 10 second interval not exceed $\sim$\,$10^{53}$ erg, so our criterion
is
\be
\Delta E_{\rm SN} \sim (V \Delta t) \frac{d\rho}{dt}   = 
\frac{ 3 \, m \Gamma }{2\pi^2}   (V \Delta t)  \int_0^\infty   d  p \, p^2         
 e^{-E_{\zp}/T}  \,  e^{-  \Gamma_{\zp} R m_{\zp}/ E_{\zp}} \lesssim  10^{53} \, {\rm erg},
\ee
where we have dropped $\zp$ subscripts and multiplied by the core volume, $V  = \frac{4}{3} \pi R^3$. 

\subsection*{Comparing Other Processes}
The $\zp$ can also be produced via $\nu$ scattering with radiative emission $\nu \nu \to \nu \nu \zp$ which has the same scaling in the gauge coupling as inverse decays, but
doesn't depend on $m_\zp$ for $m_{\zp} \ll T \sim 30$ MeV. The radiative production rate is approximately 
\be
\Gamma(\nu \nu \to \nu \nu \zp)= n_\nu \sigma(\nu \nu \to \nu \nu \zp) \sim  G_F^2 T^5 \alpha_{\mu-\tau} \sim  3.2 \times 10^{-18} \, {\rm GeV} \, \alpha_{\mu-\tau},
\ee
where we have taken $T \sim 30$ MeV, included a $ \alpha_{\mu-\tau}$ penalty for $\zp$ emission, and conservatively neglected the 3-body phase suppression for this process. 
Comparing this to the thermally averaged inverse decay process with the same parametric scaling,
\be
\langle\Gamma_\zp\rangle  =\Gamma_{\zp} \frac{K_1(z) }{ K_2(z)} \simeq \frac{1}{3}\alpha_{\mu -\tau} m_{\zp} \left( \frac{m_\zp}{2T}\right) 
\simeq 5.5 \times 10^{-12} \, {\rm GeV} \,\alpha_{\mu-\tau} \left( \frac{m_\zp}{\rm keV} \right)^2,
\ee
we find that the latter dominates over the full mass range we consider. Only near $m_\zp \sim $ eV does radiative emission compete.
Similarly, we can estimate neutrino annihilation, $\nu \nu \to \zp\zp$, for which the rate scales as $\sim \alpha_{\mu-\tau}^2 T$ and yields
\be
\frac{      \Gamma(\nu \nu \to \zp \zp)     }{\langle\Gamma_\zp\rangle} \sim \frac{6 \alpha_{\mu-\tau}  T^2}{m_\zp^2} \sim    5.4 \times 10^{-3}\left( \frac{\alpha_{\mu-\tau}}{10^{-12}}\right) \left( \frac{\rm keV}{m_\zp}\right)^2,
\ee
so unless the mass is very small or the gauge coupling is large and in the thermal regime, inverse decays are the dominant form of $\zp$ production.
Note that in this estimate, we have taken $T = 30$ MeV for the SN temperature. 

\section{Scalar Loop Induced Kinetic Mixing }\label{app:mu-tau_loop}
\renewcommand{\theequation}{C.\arabic{equation}}
\setcounter{equation}{0}

\begin{figure}[t]
\centering
\begin{tabular}{cc}
\includegraphics[width=0.4\textwidth]{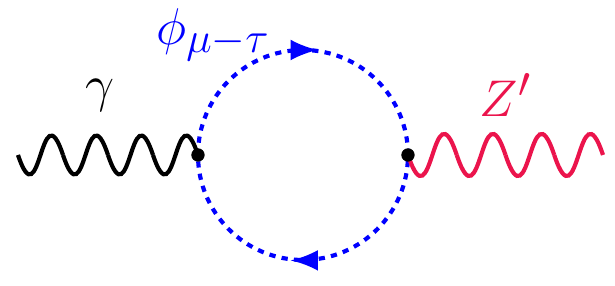} & \includegraphics[width=0.4\textwidth]{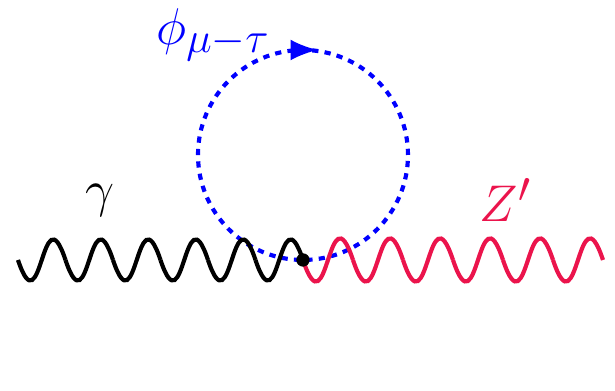} 
\end{tabular}
\caption{Contributions to the $\gamma-\zp$ kinetic mixing from heavy scalar loops.}\label{fig:diagram_scalars}
\label{fig:scalarloops}
\end{figure}

In addition to the irreducible ``natural" kinetic mixing considered throughout this paper, there could also be 
additional contributions from hypothetical heavier particles, beyond the SM. 
To assess the magnitude of such contributions, we calculate in this appendix the contribution to $\gamma-\zp$ kinetic mixing from an additional pair of heavy scalars with identical charges under electromagnetism and opposite charges under $L_\mu - L_\tau$.

The loop-level kinetic mixing from each scalar arises from two intermediate loop diagrams, as shown in Fig.~\ref{fig:scalarloops}. The two propagator contribution is 
\be
i\Pi^{\mu\nu}_2(q^2) = i^2 (ie)(ig_{\mu-\tau}) \int \frac{d^4 k}{(2\pi)^4} \frac{   (2k + q)^\mu  (2k + q)^\nu   }{[(k+q)^2 - m^2   ] [ k^2  - m^2  ]},
 \ee
where $m$ is a scalar mass and we combine denominators with Feynman parameters 
\be
\frac{1}{[(k+q)^2 - m^2   ] [ k^2  - m^2  ]} = \int_0^1 \frac{dx}{ [ \ell^2 + x(1-x) q^2 - m^2]^2 } \equiv  \int_0^1 \frac{dx}{ (\ell^2 - \Delta)^2 },
\ee
where $\ell \equiv k + xq$ and $\Delta \equiv - x(1-x)q^2 + m^2$. Dropping terms linear in $\ell$ in the numerator of the 
integrand, we can rewrite this as 
\be
i\Pi^{\mu\nu}_2(q^2) = e \gmt \int_0^1 dx    \left(   
 4 \int \frac{d^4 \ell}{(2\pi)^4} \frac{    \ell^\mu \ell^\nu   }{ ( \ell^2 - \Delta)^2 }
 +
  (1-2x)^2 q^\mu q^\nu  \int \frac{d^4 \ell}{(2\pi)^4} \frac{  1  }{ ( \ell^2 - \Delta)^2 }
      \right),
 \ee
which we can evaluate 
 in dimensional regularization and extract the divergent piece
 \be
({\rm divergent}) ~ ~  i\Pi^{\mu\nu}_2(q^2) 
&=&  \frac{  ie  \gmt  m^2 }{4 \pi^2 \epsilon}\,  g^{\mu\nu}   +     
\frac{  ie  \gmt }{24 \pi^2 \epsilon}   ( q^\mu q^\nu  -  q^2 g^{\mu\nu}  ) ,
 \ee
where the second term has the Ward identity structure, but the first term will need to cancel against the 
contribution from the one propagator diagram with a four point vertex, which is simpler to evaluate
  \be
 i \Pi^{\mu\nu}_1(q^2) = i (2ie  \gmt) g^{\mu\nu} \int \frac{d^4 k}{(2\pi)^4} \frac{1}{  k^2  - m^2  } ,
 \ee
which yields a divergent piece 
\be
({\rm divergent})  ~~~ i\Pi^{\mu\nu}_1(q^2)   =  -   \frac{ ie  \gmt m^2}{4\pi^2 \epsilon}    g^{\mu\nu}.
 \ee
Combining the divergent $\epsilon$ pole contributions now becomes
\be
({\rm divergent}) ~ ~ i \Pi^{\mu\nu}_1(q^2) +  i \Pi^{\mu\nu}_2(q^2) &=&  \frac{  ie  \gmt }{24 \pi^2 \epsilon}   ( q^\mu q^\nu  -  q^2 g^{\mu\nu}  ) ,
\ee
which has the requisite gauge invariant structure.  
  In the $\overline{\rm MS}$ scheme the residual finite contribution is
\be
i\Pi^{\mu\nu}(q^2)  &=& \frac{ie \gmt}{16 \pi^2 }  \int_0^1 dx    \biggl(   
 2 g^{\mu\nu}     \biggl[     m^2 \left(    \log m^2-1 \right)  -           \Delta  \left(   \log \Delta -1 \right)  \biggr]
 -  (1-2x)^2 q^\mu q^\nu    \log \Delta  
      \biggr),~~
\ee
so performing the integrals yields the gauge invariant result
\be
\label{eq:epsilon-qfunc}
i\Pi^{\mu\nu}(q^2)  =  \frac{ie \gmt}{144 \pi^2  q^2} \! \left[  24 m^2 - 8q^2 - 6 q^2 \left( \frac{4m^2}{q^2}  - 1 \right)^{\!\! 3/2} \! \cot^{-1} \sqrt{ \frac{4m^2}{q^2}  - 1}     + 3 q^2 \log m^2  \right] \!\! \left(  q^\mu q^\nu - q^2 g^{\mu\nu} \right).~~~~~
\ee
To eliminate the explicit dimensionful $\log m^2$ dependence, we need to compare this to a reference value, so just like with the fermion case with $\mu$ and $\tau$ loops, we add a second scalar with opposite $L_\mu - L_\tau$ charge and identical SM quantum numbers to obtain a mixing with the hypercharge gauge boson
\be
\label{eq:epsilon-qfunc-2}
i\Pi^{\mu\nu}(q^2)  =  \frac{ie \gmt}{48 \pi^2  q^2} \! \left(  8 (m_1^2 - m_2^2) - 2 q^2 (  x_1^3 \cot^{-1} x_1 -   x_2^3 \cot^{-1} x_2 )      +  q^2 \log \frac{ m_1^2 }{m_2^2} \, \right) \! \left(  q^\mu q^\nu - q^2 g^{\mu\nu} \right),~~~~~~~~
\ee
where $m_{1,2}$ are the scalar masses, $x_i \equiv \sqrt{4m_i^2/q^2  - 1}$. We can extract the kinetic mixing from the coefficient of $\left(     q^2 g^{\mu\nu} - q^\mu q^\nu \right)$ by expanding in powers
of $q^2$ to get
\be
 \epsilon \simeq  \frac{e g_{\mu-\tau}}{48 \pi^2} \log \frac{m_1^2}{m_2^2}  \sim 3.6 \times 10^{-3} g_{\mu-\tau}, 
 \ee
where we have taken the limit $q^2 \to 0$ and assumed $m_1/m_2 = m_\tau/m_\mu$ in the log.
If the scalars are integrated out above the electroweak scale, the only allowed kinetic mixing is between the $\zp$ and 
the SM hypercharge gauge boson, $B_\mu \equiv \cos \theta_W A_\mu - \sin\theta_W Z^0_\mu$, so we would need to replace $e \to g^\prime$ 
where $g^\prime \equiv e \cos\theta_W$ is the hypercharge gauge coupling.
Since the induced kinetic mixing is logarithmically sensitive to heavy particle masses, there is considerable UV dependence to the ``natural" value.

\renewcommand{\theequation}{D.\arabic{equation}}
\setcounter{equation}{0}
\section{Delayed Neutrino Decoupling}\label{app:App_nudec}

In this appendix we derive the $\zp$ mediated corrections to the energy transfer rates in Eqs.~(\ref{eq:DrhoZp}) and (\ref{eq:Drhonumu}), which
 are responsible for delaying neutrino decoupling; the SM processes which are also present and available in \cite{Escudero:2018mvt} and in references therein. The Liouville equation that evolves the phase space density $f_i$
  of particle species $i$ in an expanding FRW universe is 
 \be
 \label{eq:liouville}
 \frac{df_i}{dt} = \frac{\partial f_i}{\partial t} - H p_i \frac{\partial f_i}{\partial p_i} = C_{i}[f_i],
 \ee
 where $p_i$ is the particle's three momentum and the collision term for $i +X \to Y$ is
 \be
 C_{i}[f_i] = \frac{1}{2E_i}    \prod_{X,Y} \frac{d^3 \vec p_X}{(2\pi)^3 2 E_X} \frac{d^3 \vec p_Y}{(2\pi)^3 2 E_Y} S  |{\cal A}_{i+X\to Y}|^2 \Lambda(f_i,\{f_Y\},\{f_Z\} ) (2\pi)^4 \delta \left(p_i + \Sigma_Y \, p_Y -  \Sigma_Z \, p_Z\right),~~~~~
 \ee
 where $X$ and $Y$ are other
 initial and final state species, ${\cal A}_{i+X\to Y}$ is the matrix element for this process, $S$ is a symmetrization factor which is $1/2$ for identical particles in the initial or final states,
 and
 \be
 \Lambda(f_i,\{f_Y\},\{f_Z\} ) \equiv f_i \prod_Y f_Y \prod_Z(1 \pm f_Z) -  \prod_Z  f_Z     \prod_Y  (1 \pm f_Y)(1\pm f_i),
 \ee
 is the phase space factor where the $\pm$ signs correspond to Bose and Pauli statistics, respectively.

Dividing Eq.~(\ref{eq:liouville}) by $(2\pi)^3$ and integrating over $g_i \, d^3 \vec p_i$ gives the familiar Boltzmann equation
 for the evolution of the number density
\be
 \frac{ \partial n_i}{\partial t}  +  3 Hn_i = \int \frac{d^3\vec p_i}{(2\pi)^3 2E_i} \, g_i \, C_i[f_i]  \equiv \frac{\delta n_i}{\delta t}. 
\ee
Similarly, dividing Eq.~(\ref{eq:liouville}) by $(2\pi)^3$ and integrating over $g_i \, E_i\, d^3 \vec p_i $ we can track the evolution of the energy
 density 
\be
 \frac{ \partial \rho_i}{\partial t}  +  3 H ( \rho_i+ P_i) = \int \frac{d^3\vec p_i}{(2\pi)^3 2E_i}\, g_i \, E_i \, C_i[f_i]  \equiv \frac{\delta \rho_i}{\delta t},
\ee
where $P_i$ is the pressure and we have defined an energy transfer rate  $\delta \rho_i / \delta t$. Thus, in the context of this work, 
computing the neutrino decoupling temperature is tantamount to evaluating integrals over the collision term.

 \subsection*{Inverse Decays}
For the inverse decay process $e^+(p_1) e^-(p_2) \to \zp(k)$ in  Eq.~(\ref{eq:DrhoZp}) the energy transfer rate is 
\be
 \int  \frac{d^3 \vec k}{(2\pi)^3 2 E_k}    \frac{d^3 \vec p_1}{(2\pi)^3 2 E_1}
 \frac{d^3 \vec p_2}{(2\pi)^3 2 E_2}  g_k \, E_k |{\cal A}_{e^+e^- \to \zp}|^2 f^{\rm (eq)}_e(\vec p_1) f^{\rm (eq)}_e(\vec p_2)   (2\pi)^4 \delta^4(p_1+p_2 - k) ,~~~~~
\ee
where $|{\cal A}_{e^+e^- \to \zp}|^2$ is the inverse decay matrix element, $f^{\rm (eq)}(p)$ is approximated to be the Maxwell-Boltzmann distribution, and we have neglected Pauli blocking effects. Using energy conservation, 
detailed balance $ f^{\rm (eq)}_e(\vec p_1) f^{\rm (eq)}_e(\vec p_2) =  f^{\rm (eq)}_{\zp}(\vec k)$, and
 the definition of the partial width for $\Gamma_{\zp \to e^+e^-}$ we can simplify this expression 
\be
\frac{\delta \rho_{\zp} }{\delta t} = \frac{3 m_{\zp} \Gamma_{\zp \to e^+e^-} }{2 \pi^2 }  \int_{m_\zp}^\infty  dE E \sqrt{E^2 - m_{\zp}^2} e^{-E/T_\gamma}   = 
 \frac{3 m^3_{\zp} T_\gamma }{2 \pi^2 } K_2\left(\frac{m_{\zp}}{T_\gamma}\right)        \Gamma_{\zp \to e^+e^-},
\ee
where we have dropped the $k$ subscript on the integration variable. Note that this result recovers
 the first term on the RHS of Eq.~(\ref{eq:DrhoZp}); the reverse process with $T_\gamma \to T_{\nu_\mu}$ accounts  for the second term,
 so we have 
 \be
 \frac{ \delta \rho_{Z'}}{\delta t} & =&  \frac{3 m_{Z'}^3}{2\pi^2} \left[ T_{\gamma} K_2\left(\frac{m_{Z'}}{T_{\gamma}}\right) - T_{\nu_\mu} K_2\left(\frac{m_{Z'}}{T_{\nu_\mu}}\right) \right]   \Gamma_{Z' \to e^+ e^-} ,~
 \ee
 which recovers the expression in Eq.~(\ref{eq:DrhoZp}).

\subsection*{Annihilation and Scattering}
For the 2-2 processes in Eq.~(\ref{eq:Drhonumu}), we follow the formalism outlined in Refs.~\cite{Kawasaki:2000en,Dolgov:2002wy,Escudero:2018mvt}, which compute the collision terms for   $\bar \nu \nu \leftrightarrow e^+e^-$ annihilation and 
$\nu e \leftrightarrow \nu e$ scattering in the SM. We are interested in using these results to compute the $\zp$ mediated corrections to neutrino decoupling in our $L_\mu - L_\tau$ extension with $\zp-\gamma$ kinetic mixing. 
To simplify our analysis, we take the massless electron limit
and integrate out the $\zp$ to obtain an effective Lagrangian for $\nu_{\mu, \tau}$ and electron interactions, which contains the interactions  
\be
\label{eq:leff}
 - 2 \sqrt{2} G_F [ \bar \nu_\mu \gamma^\mu P_L \nu_\mu +  \bar \nu_\tau \gamma^\mu P_L \nu_\tau][ \bar e \gamma_\mu (C_V - C_A\gamma^5 ) e]  -
 \frac{\epsilon e g_{\mu-\tau}}{m_{\zp}^2}[ \bar \nu_\mu \gamma^\mu P_L \nu_\mu - \bar \nu_\tau \gamma^\mu P_L \nu_\tau][ \bar e \gamma_\mu e],~~~~~
\ee
where the first term represents SM interactions with vector and axial couplings $C_{V,A}$ and the second term arises from virtual $\zp$ exchange. From \cite{Kawasaki:2000en,Dolgov:2002ab}, the SM energy transfer rates for scattering and annihilation 
can be written as \cite{Escudero:2018mvt}
\be
\frac{1}{2\pi^2} \int  dp p^3 C_{\bar \nu \nu \leftrightarrow e^+e^-} &=& \frac{16 G_F^2 (C_V^2 + C_A^2) }{\pi^5} 
\left( T_\gamma^9 - T_\nu^9 \right) \, ,\\
\frac{1}{2\pi^2} \int dp p^3 C_{\bar \nu e \leftrightarrow \nu e} &=& \frac{7 G_F^2 (C_V^2 + C_A^2) }{\pi^5} T^4_\gamma T^4_\nu (T_\gamma - T_\nu),
\ee
where we have neglected all chemical potentials and included all forwards and backwards reactions.
We can use this result to compute the corresponding $\zp$ induced corrections
 by noting that the new interactions in Eq.~(\ref{eq:leff}) have the same form as their SM electroweak counterparts
with the replacements 
\be
C_V \to 1~~,~~ C_A \to 0 ~~,~~G_F \to \frac{\epsilon e g_{\mu-\tau}}{2\sqrt{2} m_{\zp}^2},
\ee
so the additional energy transfer rates in an $L_\mu - L_\tau$ extension are 
\be
\left( \frac{  {\delta \rho_\nu}}{\delta t} \right)_{\! \zp} = \frac{(e\epsilon g_{\mu-\tau})^2}{ 4\pi^5 m_{\zp}^4} 
\biggl[         16(T_\gamma^9 - T_{\nu}  ^9) + 7T_\gamma^4 T_{\nu}  ^4 (T_\gamma-T_{\nu}  )     \biggr],   
\ee
which recovers the $g_{\mu -\nu}$ dependent correction in Eq.~(\ref{eq:Drhonumu}). Note that this expression includes an additional overall factor of 2 to account for both neutrinos and antineutrinos. 
\begin{figure}[t]
\begin{tabular}{cc}
\includegraphics[width=0.47\textwidth]{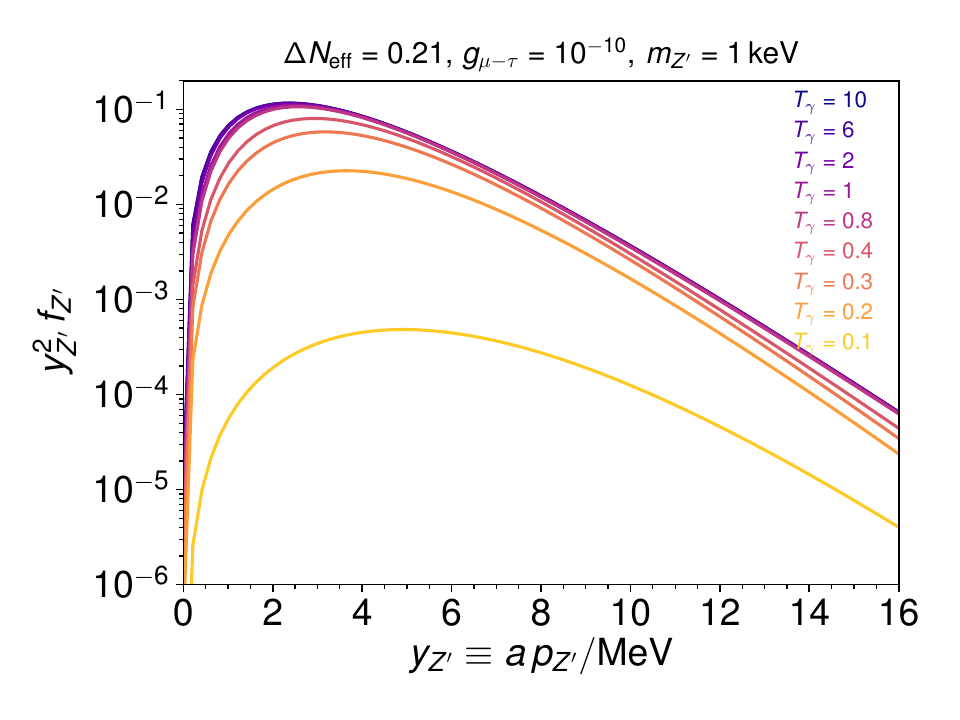}  & \includegraphics[width=0.47\textwidth]{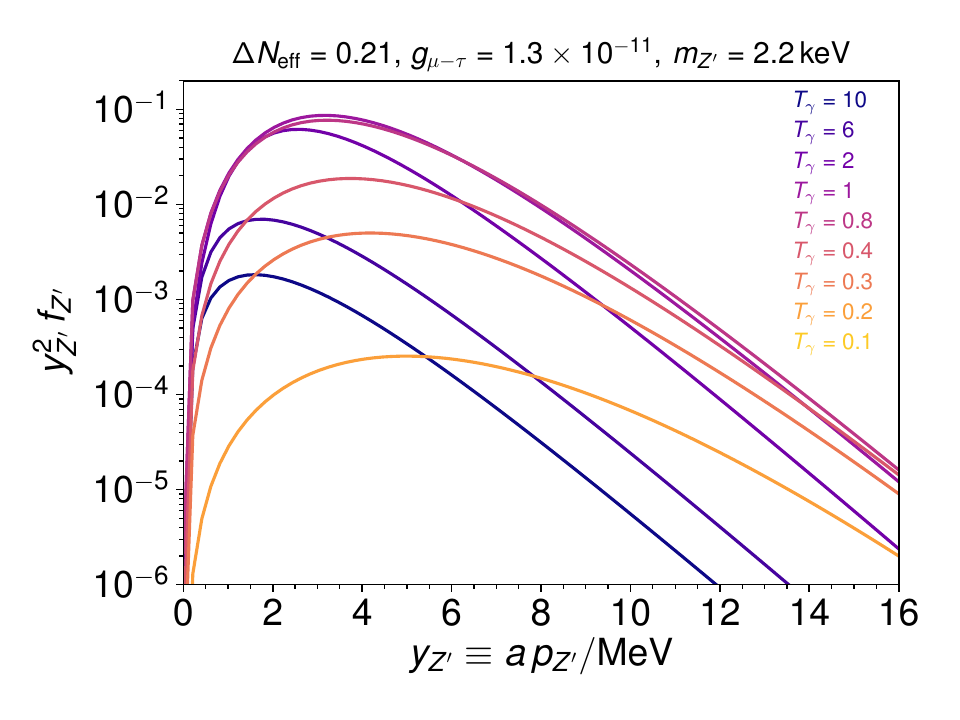} \\
\includegraphics[width=0.47\textwidth]{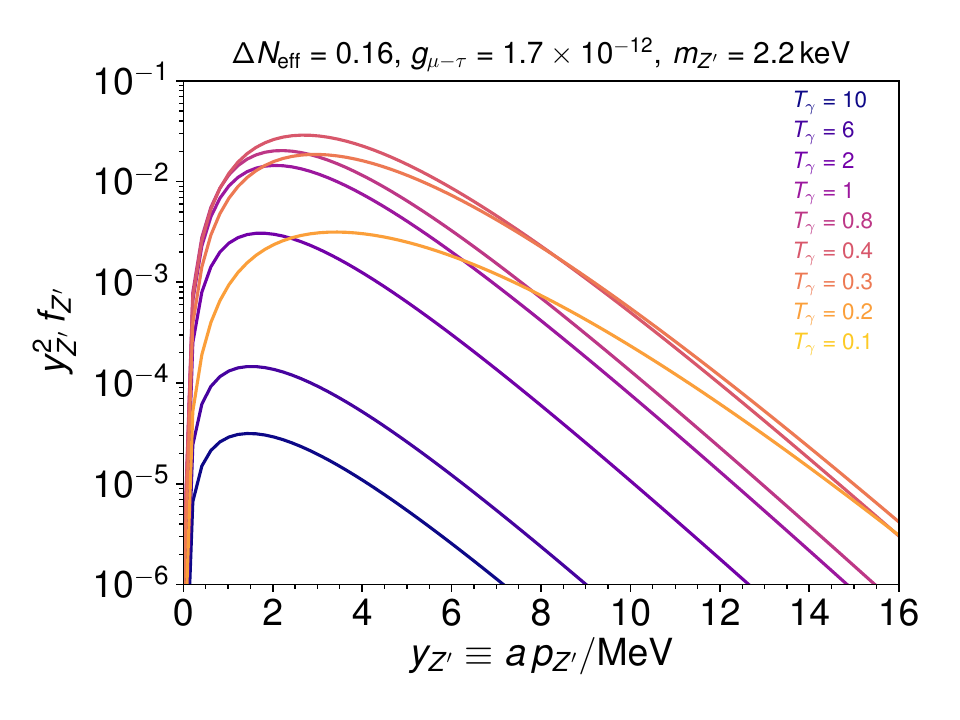}  & \includegraphics[width=0.47\textwidth]{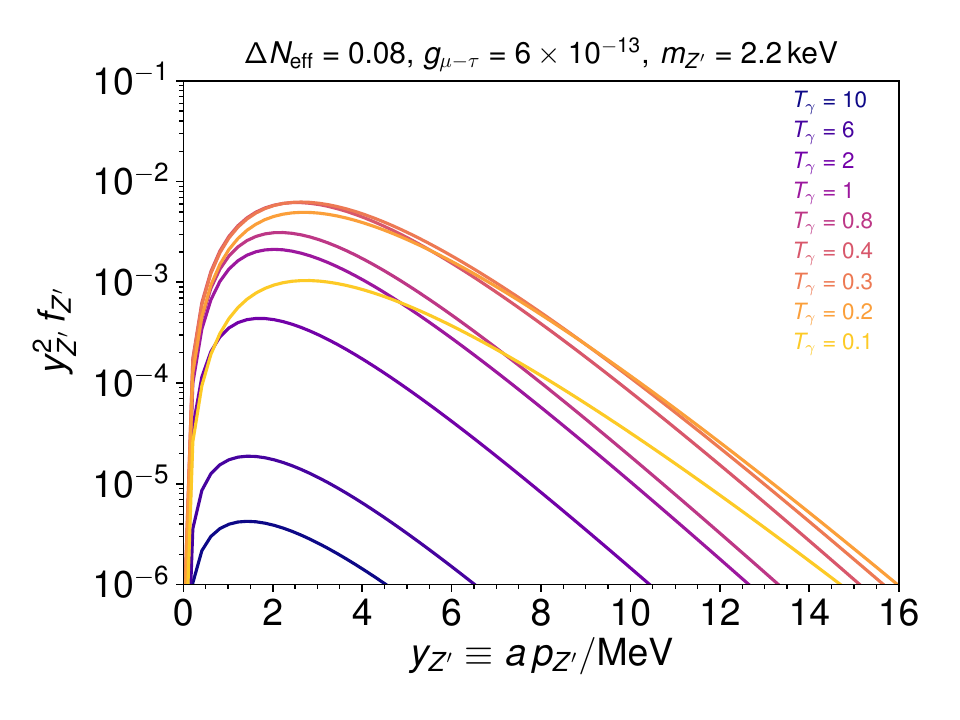} 
\end{tabular}
\caption{In the upper panels, we plot the evolution of the $Z'$ distribution function as a function of the photon temperature (in keV) for the two choices of masses and couplings considered in Fig.~\ref{fig:evZpfreezeIN} which lead to $\Delta N_{\rm eff} = 0.21$. The upper left panel corresponds to a $Z'$ that thermalizes with the neutrinos while relativistic, while in upper right panel the $Z'$ thermalizes with the neutrinos at $T_\nu \sim m_{Z'}$. In the lower panels, we plot the evolution of the $Z'$ distribution as a function of the photon temperature for two scenarios in which the $Z'$ never reaches thermal equilibrium with the neutrinos, leading to $\Delta N_{\rm eff} < 0.21$.  }
\label{fig:f_Z}
\end{figure}

\newpage
\section{Freeze-in Solutions}\label{app:App_Freeze-In}

In this Appendix, we describe some of the solutions to the time evolution of the $Z'$ and neutrino distribution functions as obtained by integrating Eqns.~\eqref{eq:ftosolve} and~\eqref{eq:ftosolve_fnu} as described in Sec.~\ref{sec:early-universe}. In the upper panels of Fig.~\ref{fig:f_Z} we show the evolution of $f_{Z'}$ as a function of the photon temperature for the two scenarios considered in Fig.~\ref{fig:evZpfreezeIN}. In the upper left panel we show the case in which the $Z'$ reaches thermal equilibrium with the neutrino population while relativistic, while the upper right panel corresponds to a scenario in which the $Z'$ thermalizes when $T_\nu \sim m_{Z'}$. The two lower panels show the evolution of the $Z'$ distribution function for two choices of $g_{\mu-\tau}$ and $m_{Z'}$ for which the $Z'$ never reaches thermal equilibrium with the neutrinos, leading to $\Delta N_{\rm eff} < 0.21$. 

\begin{figure}[t]
\centering\hspace{-0.5cm}
\begin{tabular}{cc}
\includegraphics[width=0.6\textwidth]{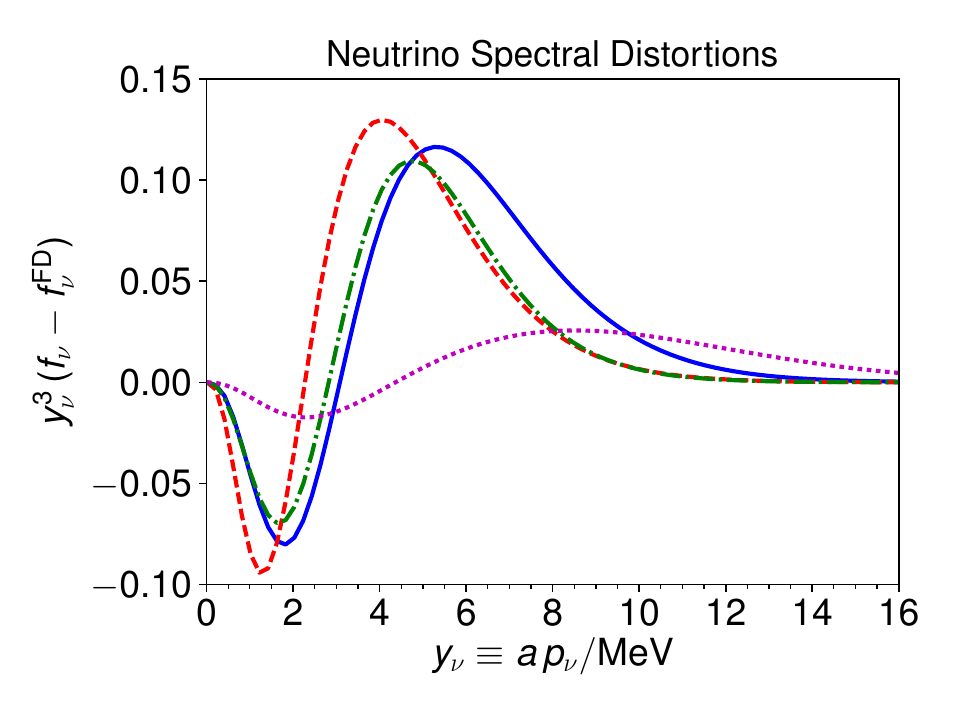}  &\includegraphics[ height=0.35\textheight]{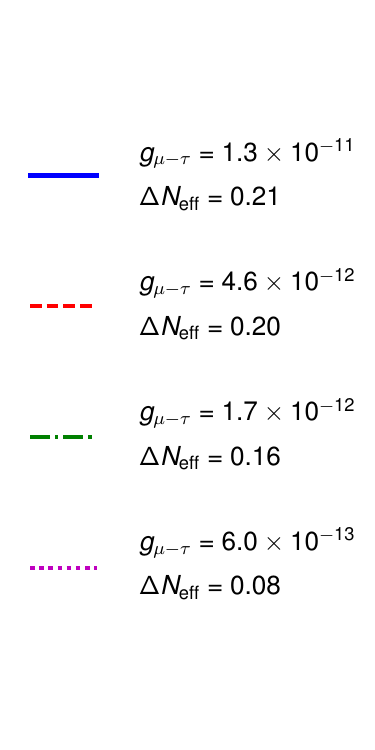}
\end{tabular}

\caption{The resulting neutrino distribution functions, $f_\nu$, after the $Z'$ population has completely decayed as a function of the comoving momentum, $y_\nu$. $f_\nu^{\rm FD}$ is the distribution function for a free-streaming decoupled neutrino, $f_\nu^{\rm FD} = 1/(1+e^{y_\nu})$. Here we have chosen $m_{\zp} = 2.2\,\,\text{keV}$.}
\label{fig:f_nu}
\end{figure}

In Fig.~\ref{fig:f_nu} we show the neutrino distribution function after a $\zp$ population that was initially generated through $\bar{\nu} \nu \to Z'$ inverse decays has complete decayed away (at $T_\nu \ll m_{Z'}$, as relevant for CMB observations). The $y$ axis has been normalized such that $\Delta N_{\rm eff}$ can be computed as $\Delta N_{\rm eff} = 3 \times \frac{120}{7\pi^4} \int_0 ^\infty\, dy\, y^3 (f_{\nu}-f_\nu^{\rm FD})$, where $f_\nu^{\rm FD}$ corresponds to the distribution function of a free-streaming decoupled neutrino, $f_\nu^{\rm FD} = 1/(1+e^{y_\nu})$. Note that the results for a different $Z'$ mass can be obtained by rescaling the corresponding couplings in Fig.~\ref{fig:f_nu} by $ (2.2\,\text{keV}/m_{Z'})^{-1/2} $.

Fig.~\ref{fig:f_nu} illustrates why a $Z'$ that thermalizes with the neutrinos after they have decoupled from the SM plasma renders $\Delta N_{\rm eff} \simeq 0.21$. Once the neutrinos have decoupled from the SM plasma, the only relevant process are $\bar{\nu}\nu \to  Z'$ and $Z' \to \bar{\nu}\nu$. At very high temperatures $f_{Z'} = 0$, and a $Z'$ population is eventually generated at the expense of neutrinos, via $\bar{\nu}\nu \to Z'$ (this can also be seen from Fig.~\ref{fig:evZpfreezeIN}). After the $Z'$ population has thermalized with the neutrinos, it decays out of equilibrium to neutrinos. However, the decay products of the $Z'$ population have an energy $E_\nu \sim m_{Z'}/2$, which is substantially different from $3\, T_\nu$, leading to a final neutrino population that is more energetic than that found in thermal equilibrium (as can be appreciated from Fig.~\ref{fig:f_nu}). This results in $\Delta N_{\rm eff} \simeq 0.21$ for $g_{\mu-\tau} \gsim 1.3\times 10^{-10}\, (m_{Z'}/\text{MeV})^{1/2}$, and to $\Delta N_{\rm eff} < 0.21$ for smaller values of $g_{\mu-\tau}$ (which do not enable the $Z'$ population to reach thermal equilibrium with the neutrinos).

\end{document}